\documentclass[submission, Phys]{SciPost}
\usepackage{scalerel,amssymb}
\usepackage[force]{feynmp-auto}	
\usepackage{slashed}		
\usepackage{xcolor}			
\usepackage{bbm}			
\usepackage{braket}	
\usepackage{dsfont}			
\usepackage{hyperref}		
\usepackage{lscape}
\usepackage{pdflscape}
\usepackage{wasysym}

\definecolor{dgreen}{HTML}{008000}

\definecolor{dblue}{HTML}{0000A0}

\definecolor{nicered}{rgb}{0.7,0.1,0.1}
\definecolor{nicegreen}{rgb}{0.1,0.5,0.1}
\definecolor{niceblue}{rgb}{0.0,0.1,0.7}
\hypersetup{colorlinks,citecolor=nicegreen,linkcolor=nicered,urlcolor=niceblue}

\def \bm#1{\mbox{\boldmath$#1$\unboldmath}}
\def \beq{\begin{equation}}
\def \eeq{\end{equation}}
\def \bea{\begin{eqnarray}}
\def \eea{\end{eqnarray}}
 
\usepackage{mathtools}

\usepackage{subcaption}
\usepackage{pgfplots}
\pgfplotsset{compat=1.16}
\usepgfplotslibrary{colormaps}

\begin{document}
\begin{fmffile}{main}
\fmfset{arrow_len}{3mm}

\begin{flushright}
 MPP-2024-236
 \end{flushright}
 
 \vspace{2mm}

\begin{center}{\Large \textbf{A tale of $\bm{Z}$+jet: SMEFT effects and the Lam-Tung relation}}
\end{center}

\begin{center}
R. Gauld,\textsuperscript{1} U. Haisch,\textsuperscript{1} and J.~Weiss\textsuperscript{1,2}
\end{center}

\begin{center}
{\bf 1} Werner-Heisenberg-Institut, Max-Planck-Institut für Physik, \\ Boltzmannstraße 8, 85748 Garching, Germany \\[2mm]
{\bf 2} Technische Universit{\"a}t M{\"u}nchen, Physik-Department, \\ James-Franck-Straße 1, 85748 Garching, Germany \\[2mm]
\href{mailto:rgauld@mpp.mpg.de}{rgauld@mpp.mpg.de}, 
\href{mailto:haisch@mpp.mpg.de}{haisch@mpp.mpg.de}, 
\href{mailto:jweiss@mpp.mpg.de}{jweiss@mpp.mpg.de}
\end{center}

\begin{center}
\today
\end{center}


\section*{Abstract} {\bf We derive constraints on dimension-six light-quark dipole operators within the Standard Model (SM) effective field theory, based on measurements of 
$\bm{Z}$ production at SLC~and~LEP, as well as $\bm{Z}$+jet production at the LHC. Our new constraints exclude the parameter space that could potentially explain the observed discrepancy between theoretical predictions and experimental data for the Lam-Tung relation. With these updated limits, we model-independently determine the maximum possible influence that beyond-SM contributions could have on the angular coefficients $\bm{A_0}$ and $\bm{A_2}$, which enter the Lam-Tung relation.}

\vspace{10pt}
\noindent\rule{\textwidth}{1pt}
\tableofcontents\thispagestyle{fancy}
\noindent\rule{\textwidth}{1pt}
\vspace{10pt}

\section{Motivation}
\label{psec:introduction}

Precision measurements at the Large Hadron Collider~(LHC) are key to testing the Standard Model~(SM). Processes with a distinct and clean signature, such as the decay of a $Z$~boson into two charged leptons, are thereby of particular interest. In fact, the transverse momentum~($p_T$) distribution of the $Z$ boson in neutral-current (NC) Drell-Yan~(DY) production is one of the most precisely measured and predicted observables at the~LHC, with an experimental accuracy of better than $1\%$ for $p_{T,Z}$ values below $250 \, {\rm GeV}$~\cite{ATLAS:2019zci,CMS:2022ubq}. These measurements are pivotal for the high-precision determinations of both the strong coupling constant $\alpha_s$~\cite{ATLAS:2023lhg} and the $W$-boson mass~\cite{ATLAS:2024erm,CMSMW}, to highlight just two of the most prominent LHC applications within the SM. 

In addition to offering important tests of the SM, the~$p_{T,Z}$ spectrum can also reveal potential signs of beyond SM~(BSM) phenomena. Within the SM, the transverse momentum of the $Z$~boson is mainly generated at lower~$p_T$, driven by the recoil of both soft and hard initial-state collinear QCD radiation. However, in the high transverse momentum region, deviations from the SM predictions may arise due to the presence of new particles or interactions not included in the SM. Such scenarios, like the existence of heavy new particles (e.g.,~extra gauge bosons or dark matter candidates), can result in an excess of DY~events with high~$p_T$. Under the assumption that these new degrees of freedom cannot be directly produced, deviations of this type can be interpreted in a largely model-independent manner using an effective field theory~(EFT) approach, such as the SMEFT~\cite{Buchmuller:1985jz,Grzadkowski:2010es,Brivio:2017vri,Isidori:2023pyp}. Due to their significant phenomenological importance, studies of DY~processes have become a central focus of the~LHC's SMEFT program --- see, for example, the publications~\cite{Alioli:2018ljm,Dawson:2018dxp,daSilvaAlmeida:2019cbr,Alioli:2020kez,Horne:2020pot,Torre:2020aiz,Greljo:2021kvv,Panico:2021vav,Dawson:2021ofa,Boughezal:2022nof,Allwicher:2022mcg,Boughezal:2023nhe,Li:2024iyj}. 

Most SMEFT studies on DY~production have focused on the impact of four-fermion contact interactions at tree level or loop-level effects from gauge-boson operators, with fewer analyses dedicated to dipole operators~\cite{daSilvaAlmeida:2019cbr,Li:2024iyj}. Previous studies have placed bounds on light-quark dipole couplings through electroweak~(EW) precision measurements at the $Z$-pole and analyses of EW diboson and DY production at the LHC~\cite{daSilvaAlmeida:2019cbr}. Additionally, it has been proposed that this form of BSM physics could help resolve potential discrepancies between theoretical predictions and experimental data for the angular coefficients $A_0$~and~$A_2$~\cite{Li:2024iyj}, whose difference provides a test of the Lam-Tung relation~\cite{Lam:1978zr,Lam:1978pu,Lam:1980uc}. Within the~SM, this relation, valid to ${\cal{O}}(\alpha_s)$, implies $A_0 = A_2$. The main aim of this article is to reassess the findings of~\cite{daSilvaAlmeida:2019cbr,Li:2024iyj} by utilizing SLC~and~LEP data on $Z$~production, alongside LHC data on $Z$+jet production. Specifically, we obtain updated constraints on light-quark dipole interactions using precision measurements of the partial $Z$-boson decay widths in $e^+e^-$ collisions and the normalized $p_{T,Z}$ spectrum in NC~DY production in~$pp$~collisions. With these updated constraints, we evaluate the maximum possible violation of the Lam-Tung relation that such BSM effects could induce. Our~results indicate that light-quark dipole operators cannot account for the aforementioned discrepancy.

This work is organised as follows: in~Section~\ref{sec:setup} we detail the theoretical ingredients that are relevant in the context of this article. Our~discussion covers the structure of light-quark dipole interactions within the~SMEFT framework, an analysis of the BSM modifications of the partial $Z$-boson decay widths and the corresponding $Z$+jet matrix elements, and a concise review of the Lam-Tung relation. The experimental data and the~Monte~Carlo~(MC) setup used to generate the relevant predictions for $Z$+jet production are detailed in~Section~\ref{sec:MCsetup}. Our numerical results are presented in~Section~\ref{sec:numerics}. We~derive constraints on light-quark dipole operators using SLC~and~LEP measurements of EW~precision observables, along with current LHC data on NC~DY production. The~obtained limits are subsequently used to assess the maximum potential impact of this type of SMEFT contributions on the Lam-Tung relation. Section~\ref{sec:conclusions} summarizes our key findings and offers an outlook. This article concludes with a series of appendices. Appendix~\ref{app:ZJSM} contains a comprehensive analysis of the theoretical uncertainties in the~$p_{T,Z}$~spectrum in $Z$+jet production within the SM, while Appendix~\ref{app:fitdetails} provides details on the extraction of the Wilson coefficients of the light-quark dipole operators from our fit to the~$Z$+jet~process. Appendix~\ref{app:vectoroperator} offers a concise discussion of another type of SMEFT contributions to $Z$+jet production, while in~Appendix~\ref{app:more} we present results for $Z$+jet production using the choice of Wilson coefficient for the up-quark dipole operator as adopted in~\cite{Li:2024iyj}.

\section{Theoretical considerations}
\label{sec:setup}

In this section, we describe the different theoretical ingredients that are relevant in the context of this article. We start by revisiting the structure of light-quark dipole interactions within the~SMEFT framework. Following that, we discuss the modifications of the partial $Z$-boson decay widths, the kinetic properties of the associated $Z$+jet matrix elements, and then review the basics of the Lam-Tung relation. 

\subsection*{Light-quark dipole operators}
\label{sec:SMEFToperators}

In the Warsaw operator basis~\cite{Grzadkowski:2010es}, the dimension-six dipole interactions in the~SMEFT, involving light-quark fields and EW field strength tensors, are expressed as:
\beq \label{eq:LSMEFT}
\begin{split}
\mathcal{L}_{\rm SMEFT} & \supset \frac{C_{uB}}{\Lambda^2} \hspace{0.5mm} \Bar{Q}_L \sigma^{\mu \nu} \hspace{0.25mm} u_R \hspace{0.25mm} \widetilde{H} B_{\mu \nu} + \frac{C_{uW}}{\Lambda^2} \hspace{0.5mm} \Bar{Q}_L \sigma^{\mu \nu} \hspace{0.25mm} \sigma^a u_R \hspace{0.25mm} \widetilde{H} W^a_{\mu \nu} \\[2mm]
& \phantom{xx} + \frac{C_{dB}}{\Lambda^2} \hspace{0.5mm} \Bar{Q}_L \sigma^{\mu \nu} \hspace{0.25mm} d_R \hspace{0.25mm} H B_{\mu \nu} + \frac{C_{dW}}{\Lambda^2} \hspace{0.5mm} \Bar{Q}_L \sigma^{\mu \nu} \hspace{0.25mm} \sigma^a d_R \hspace{0.25mm} H W^a_{\mu \nu} + {\rm h.c.} 
\end{split}
\eeq
Here, $B_{\mu \nu}$ and $W^a_{\mu \nu} $ denote the $U(1)_Y$ and $SU(2)_L$ field strength tensors, respectively, and $\sigma^a$ are the Pauli matrices. We introduced $\sigma^{\mu \nu} = i/2 \left (\gamma^{\mu} \gamma^{\nu} - \gamma^{\nu} \gamma^{\mu} \right )$, with $\gamma^\mu$ being the Dirac matrices. The symbol $Q_L = (u_L, d_L)^T$ denotes the left-handed first-generation quark $SU(2)_L$ doublet, while $u_R$ and $d_R$ are the right-handed up-quark and down-quark~$SU(2)_L$ singlets. $H$ represents the SM Higgs doublet, and the shorthand notation $\widetilde H_i = \epsilon_{ij} \left (H_j \right)^\ast$, where $\epsilon_{ij}$ is totally antisymmetric with $\epsilon_{12} = 1$, is used. Finally, $\Lambda$ represents the common mass scale that suppresses all operators~in~(\ref{eq:LSMEFT}), making their Wilson coefficients $C_{qB}$ and~$C_{qW}$ for $q = u,d$ dimensionless. 

After EW symmetry breaking, the light-quark dipole interactions introduced in~(\ref{eq:LSMEFT}) take the following form
\beq \label{eq:LEFT}
\begin{split}
\mathcal{L}_{\rm LEFT} & \supset \frac{v}{\sqrt{2} \hspace{0.125mm} \Lambda^2} \sum_{q = u, d} \left ( C_{q\gamma} \hspace{0.5mm} \Bar{q}_L \sigma^{\mu \nu} \hspace{0.25mm} q_R \hspace{0.25mm} F_{\mu \nu} + C_{qZ} \hspace{0.5mm} \Bar{q}_L \sigma^{\mu \nu} \hspace{0.25mm} q_R \hspace{0.25mm} Z_{\mu \nu} \right ) + {\rm h.c.} \,,
\end{split}
\eeq
in the low-energy EFT~(LEFT). Here, $v \simeq 246 \, {\rm GeV}$ is the Higgs vacuum expectation value, while $F_{\mu \nu}$ and $Z_{\mu \nu}$ denote the field strength tensors of the photon and $Z$-boson fields, respectively. The~new~Wilson coefficients $C_{q\gamma}$ and $C_{qZ}$ are given by the following linear combinations of the original Wilson coefficients $C_{qB}$ and $C_{qW}$: 
\beq \label{eq:newwilson}
\begin{aligned}
C_{u\gamma} & = c_w \hspace{0.25mm} C_{uB} + s_w \hspace{0.25mm} C_{uW} \,, \qquad & C_{d\gamma} & = c_w \hspace{0.25mm} C_{dB} - s_w \hspace{0.25mm} C_{dW} \,, \\[2mm]
C_{uZ} & = -s_w \hspace{0.25mm} C_{uB} + c_w \hspace{0.25mm} C_{uW} \,, \qquad & C_{dZ} & = -s_w \hspace{0.25mm} C_{dB} - c_w \hspace{0.25mm} C_{dW} \,.
\end{aligned}
\eeq
The sine and cosine of the weak mixing angle are abbreviated by $s_w \simeq 0.48$ and $c_w \simeq 0.88$, respectively. 

Before proceeding, we note that the photonic dipole interactions in~(\ref{eq:LEFT}) are in general more strongly constrained than their $Z$-boson counterparts. The tightest constraints come from CP-violating observables, such as the neutron electric dipole moment. For instance, the article~\cite{Kley:2021yhn} reports the following upper bounds 
\beq \label{eq:nEDMbounds} 
\frac{\left | {\rm Im} \hspace{0.5mm} C_{u\gamma} \right |}{\Lambda^2} < \frac{1}{\left ( 64 \, {\rm TeV} \right )^2} \,, \qquad \frac{\left | {\rm Im} \hspace{0.5mm} C_{d\gamma} \right |}{\Lambda^2} < \frac{1}{\left ( 185 \, {\rm TeV} \right )^2} \,, 
\eeq
on the magnitudes of the imaginary parts of the Wilson coefficients~in~(\ref{eq:newwilson}) involving a photon normalized by two powers of the new-physics scale $\Lambda$. Measurements of the magnetic dipole moments of the neutron and proton~\cite{ParticleDataGroup:2024cfk} place bounds on the real parts of $C_{q\gamma}/\Lambda^2$. However, the resulting limits are both weaker and theoretically less reliable than those provided in~(\ref{eq:nEDMbounds}). 

To circumvent the stringent bounds on the Wilson coefficients $C_{q\gamma}$, we consider the following choices of~$C_{qB}$ and~$C_{qW}$ below: 
\beq \label{eq:killphoton}
C_u = C_{uB} = - \frac{s_w}{c_w} \hspace{0.5mm} C_{uW} \,, \qquad C_d = C_{dB} = \frac{s_w}{c_w} \hspace{0.5mm} C_{dW} \,. 
\eeq
For these choices, one has 
\beq \label{eq:CqgammaCqZ}
C_{q\gamma} = 0 \,, \qquad C_{qZ} = \frac{1}{s_w} \hspace{0.5mm} C_q \,.
\eeq
and, as a result, the~SMEFT modifications of interest in $Z$+jet production can be parameterized by the two coefficients $C_u$ and $C_d$. In the following, we will present all our results in terms of these coefficients. Notice that focusing on the invariant mass region around the $Z$-pole, i.e., $m_{ll} \simeq M_Z$, allows the photon corrections associated with $C_{q\gamma}$ to be neglected in the differential cross section for $Z$+jet production with very high precision. The same holds true for all other observables discussed below. Our BSM predictions are thus largely unaffected by the specific choice of $C_{q\gamma}$ made in~(\ref{eq:CqgammaCqZ}).

\subsection*{Partial $\bm{Z}$-boson decay widths}
\label{sec:zwidths}

To derive limits on the light-quark dipole interactions introduced in~(\ref{eq:LSMEFT}) from the EW~precision measurements at SLC~and~LEP~\cite{ALEPH:2005ab}, we examine the partial $Z$-boson decay widths. For the specific Wilson coefficients~$C_q$ in~(\ref{eq:killphoton}), and treating the light quarks as massless, we find the following modifications
\beq \label{eq:Zwidths}
\frac{\Gamma \left ( Z \to q \bar q \right)}{\Gamma \left ( Z \to q \bar q \right)_{\rm SM}} = 1 + N_q \hspace{0.25mm} \frac{v^2 M_Z^2}{\Lambda^4} \hspace{0.25mm} \left | C_q \right | ^2 \,,
\eeq
where $M_Z \simeq 91.2 \, {\rm GeV}$ denotes the mass of the $Z$ boson. The factor $N_q$ is flavor dependent and given by 
\beq \label{eq:Nq}
N_q = \frac{2}{s_w^2} \hspace{0.25mm} \frac{1}{g_{Lq}^2 + g_{Rq}^2} \,,
\eeq
with $g_{Lq} = g/c_w \left ( T^3_q - Q_q \hspace{0.25mm} s_w^2 \right )$ and $g_{Rq} = -g/c_w \hspace{0.5mm} Q_q \hspace{0.25mm} s_w^2$ representing the left-handed and right-handed couplings of the $Z$ boson to quarks of the flavor $q$, respectively --- $g$~is the $SU(2)_L$ coupling, whereas $T^3_q$ and $Q_q$ correspond to the third component of the weak isospin and the electromagnetic charge. Explicitly, one has
\beq \label{eq:NuNd} 
N_u = \frac{18 \hspace{0.25mm} c_w^2}{\pi \alpha \left ( 9 - 24 s_w^2 + 32 s_w^4 \right )} \,, \qquad 
N_d = \frac{18 \hspace{0.25mm} c_w^2}{\pi \alpha \left ( 9 - 12 s_w^2 + 8 s_w^4 \right )} \,. 
\eeq
Here $\alpha \simeq 1/128$ denotes the electromagnetic coupling constant. Note that terms linear in~$C_q$ are absent in the expression~(\ref{eq:Zwidths}) because they would be proportional to the light-quark masses $m_q$, which vanish under our assumption. In practice, treating the up and down quarks as massless is an excellent numerical approximation. As~a~result, the terms quadratic in $\left | C_q \right |$ always provide the dominant contribution to the $Z \to q \bar q$ decay when considering the effects of light-quark dipole operators.

\subsection*{Matrix elements for $\bm{Z}$+jet production}
\label{sec:matrixelements}

\begin{figure}[t!]
\begin{center}
\includegraphics[width=0.45\textwidth]{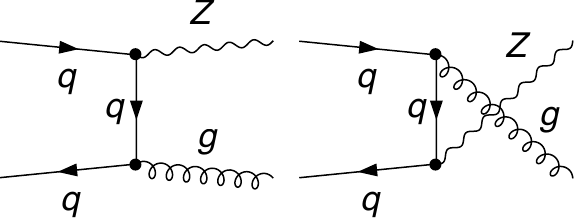} \qquad 
\includegraphics[width=0.45\textwidth]{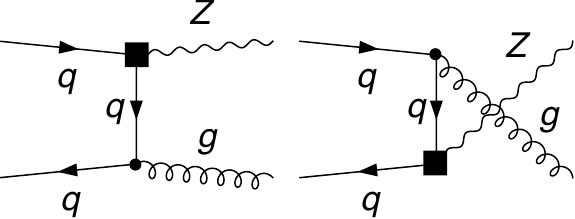}
\end{center}
\vspace{0mm} 
\caption{\label{fig:Zgdiagrams} Contributions to the $q \bar q \to Zg$ process in the SM~(left) and the~SMEFT~(right). The black squares represent an insertion of a light-quark dipole operator involving a $Z$~boson, as described by the Lagrangian~(\ref{eq:LEFT}).}
\end{figure}

In~Section~\ref{sec:numerics}, we will derive the constraints that current LHC data on NC~DY production impose on the Wilson coefficients of light-quark dipole operators involving a $Z$~boson. While those constraints are derived for the off-shell process $pp \to \gamma^\ast/Z + X \to l^+ l^- + X$, to qualitatively understand those numerical results it is useful to study the analytic structure of the on-shell Born-level matrix elements for the $Z$+jet process. To achieve this, we define the following ratios:
\beq \label{eq:defchiq}
\chi_q = \frac{\big | {\cal M}_{\rm SM} (q \bar q \to Zg) + {\cal M}_{\rm SMEFT} (q \bar q \to Zg) \big |^2}{\big | {\cal M}_{\rm SM} (q \bar q \to Zg) \big |^2} \,. 
\eeq
These ratios describe the impact of a non-zero Wilson coefficients $C_q$ in $q \bar q \to Z g$ scattering relative to the SM contributions. Relevant diagrams contributing to~(\ref{eq:defchiq}) are shown in~Figure~\ref{fig:Zgdiagrams}. 

Assuming the light quarks are massless, a straightforward tree-level calculation yields the result
\beq \label{eq:chiq}
\chi_q = 1 + N_q \hspace{0.5mm} \frac{v^2 M_Z^2}{\Lambda^4} \hspace{0.5mm} \kappa (s, t) \hspace{0.5mm} \left | C_q \right |^2 \,, 
\eeq
for the ratios defined in~(\ref{eq:defchiq}). The analytical expressions for the factor $N_q$ are provided in~(\ref{eq:Nq}) and~(\ref{eq:NuNd}), and $s = \hat s/M_Z^2$ and $t = \hat t/M_Z^2$ are the usual Mandelstam variables, normalized to the square of the $Z$-boson mass. The kinematic factor appearing in~(\ref{eq:chiq}) is instead universal and takes the following form:
\beq \label{eq:kinematic}
\kappa (s, t) = \frac{s^2 - 4 \left ( s - 1 \right )^2 t - 4 \left ( s - 1 \right ) t^2+1}{s^2 + 2 \hspace{0.125mm} s \hspace{0.125mm} t + 2 \left ( t - 1 \right ) t + 1} \,.
\eeq

A few remarks regarding~(\ref{eq:defchiq}) are in order. First, as in~(\ref{eq:Zwidths}), terms linear in~$C_q$ are absent in the expression for $\chi_q$, since these interference terms are suppressed by the small masses $m_q$, due to the chirality-flipping nature of the light-quark dipole operators. Second, in the limit of $s \to \infty$, the kinematic factor~(\ref{eq:kinematic}) behaves as 
\beq \label{eq:kappalarges}
\lim_{s \to \infty} \kappa (s,t) = \frac{1 - \cos ^2 \hat \theta}{1 + \cos ^2 \hat \theta} \hspace{0.25mm} \frac{2 \hspace{0.25mm} \hat s}{M_Z^2} \,, 
\eeq
where $\hat \theta$ represents the scattering angle between the quark and the $Z$ boson in the center-of-mass~(CM) frame. The result in~(\ref{eq:kappalarges}) demonstrates that the light-quark dipole contributions to DY~production are enhanced at high energies compared to the SM background. Similar observations have been made and utilized, for example, in~\cite{Li:2024iyj,Farina:2016rws,Greljo:2017vvb,Alioli:2017jdo,Alioli:2017nzr,Banerjee:2018bio,Grojean:2018dqj,DiLuzio:2018jwd,Dawson:2018dxp,Fuentes-Martin:2020lea,Haisch:2021hcg,Haisch:2023upo,Gauld:2023gtb,Hiller:2024vtr}. The~observed energy enhancement implies that less precise measurements of $p_{T,Z}$ at the~LHC can, in principle, achieve similar or even greater sensitivity to $C_q/\Lambda^2$ compared to the high-precision measurements of $Z \to q \bar q$ performed at SLC~and~LEP. Lastly, note that although we have focused on the $q \bar q \to Zg$ channel, the same reasoning applies to the processes $q g \to Z q$ and $\bar q g \to Z \bar q$, which can be derived from $q \bar q \to Z g$ through crossing~symmetries.

We also note that light-quark dipole operators also affect the predictions for differential NC deep inelastic scattering (DIS) measurements, including those conducted at~HERA~\cite{H1:2009pze}. A simple tree-level calculation shows that the light-quark dipole operators in question give rise to a longitudinal structure function associated with the absorption of a longitudinally polarized virtual $Z$ boson. This results in a violation of the Callan-Gross relation~\cite{Callan:1969uq} --- for an analysis of the effects of quark-lepton contact interactions on the Callan-Gross relation, see~\cite{Buchmuller:1987ur}. Although these effects are more pronounced for $\hat t \to -\infty$, the limited range of four-momentum transfer and experimental uncertainties make current NC DIS measurements inadequate for deriving competitive constraints on the Wilson coefficients of light-quark dipole operators involving a $Z$ boson.

Before proceeding, we stress that the energy enhancement of the on-shell Born-level matrix elements in $Z$+jet production, as observed in~(\ref{eq:kappalarges}), is a distinctive characteristic of the light-quark dipole operators~(\ref{eq:LSMEFT}) among the dimension-six SMEFT operators that directly alter the interactions between up and down quarks and the EW gauge bosons. This claim is illustrated in~Appendix~\ref{app:vectoroperator}, where we briefly discuss the impact of an operator that modifies the interactions between the $Z$ boson and up quarks on the $Z$+jet predictions.

\subsection*{Lam-Tung relation and its violation}
\label{sec:lamtung}

The Lam-Tung relation is a theoretical prediction related to the angular distribution of dileptons produced in the NC~DY~process, i.e., $pp \to \gamma^\ast/Z + X \to l^+ l^- + X$. In the so-called Collins-Soper (CS) frame~\cite{Collins:1977iv}, the angular distribution of the outgoing leptons can be described in terms of a set of angular coefficients~$A_i$ for $i=0,\ldots, 7$. These frame-dependent angular coefficients provide a detailed description of the lepton angular distributions as functions of the momenta of the exchanged gauge boson. Using these angular coefficients, the differential cross section can be expressed as:
\bea \label{eq:angular} 
\begin{split}
\frac{d \sigma}{dp_{T,ll} \hspace{0.25mm} dy_{ll} \hspace{0.25mm} dm_{ll}^2 \hspace{0.5mm} d \Omega} & = \frac{3}{16 \pi} \frac{d\sigma}{ dp_{T,ll} \hspace{0.25mm} dy_{ll} \hspace{0.25mm} dm_{ll}^2} \, \Bigg [ \left( 1 + \cos^2 \theta \right) + \frac{A_0}{2} \left( 1 - 3 \cos^2 \theta \right) \\[2mm]
& \phantom{xx} + A_1 \hspace{0.5mm} \sin 2\theta \cos \phi + \frac{A_2}{2} \hspace{0.5mm} \sin^2 \theta \cos 2 \phi + A_3 \hspace{0.5mm} \sin \theta \cos \phi \\[1mm]
& \phantom{xx} + A_4 \hspace{0.5mm} \cos \theta + A_5 \hspace{0.5mm} \sin^2 \theta \sin 2 \phi + A_6 \hspace{0.5mm} \sin 2 \theta \sin \phi + A_7 \hspace{0.5mm} \sin \theta \sin \phi \Bigg ] \, .
\end{split}
\eea
Here, $d \Omega = d\hspace{-0.35mm} \cos \theta \hspace{0.25mm} d \phi$ with $\theta$ and $\phi$ denoting the polar and azimuthal angles of the negatively charged lepton in the CS frame, while $p_{T,ll}$, $y_{ll}$, and $m_{ll}$ represent the transverse momentum, rapidity, and invariant mass of the lepton pair, respectively. It should be noted that~$p_{T,ll}$ is a good proxy for $p_{T,Z}$ when events are limited to the dilepton invariant mass region near $m_{ll} \simeq M_Z$, as is typically done in experimental studies. In terms of the angular coefficients~$A_i$ introduced in~(\ref{eq:angular}), the Lam-Tung relation reads: 
\beq \label{eq:lamtung}
A_0 - A_2 = 0 \,.
\eeq
This relation holds up to ${\cal O} (\alpha_s)$ in perturbative QCD under the leading-twist approximation. It arises from the fact that the NC~DY~process involves at leading-order~(LO) in QCD the annihilation of spin-1/2~quarks and antiquarks, and is further preserved at next-to-leading order~(NLO) in QCD due to the purely vectorially coupling of the spin-1~gluon to the quark current~\cite{Lam:1978zr,Lam:1978pu,Lam:1980uc,Arteaga-Romero:1983llb}. In the SM, the equality~(\ref{eq:lamtung}) is first violated in perturbation theory by ${\cal O} (\alpha_s^2)$ corrections~\cite{Mirkes:1994dp}, though the breaking of the Lam-Tung relation remains relatively small. The relation~(\ref{eq:lamtung}) therefore provides a clear prediction for the behavior of the angular coefficients $A_0$ and $A_2$ in the NC~DY~process, and its experimental study~can help in understanding the intricate dynamics that triggers the $pp \to \gamma^\ast/Z + X \to l^+ l^- + X$ process both in the SM and beyond~it.

The effects of various dimension-six and dimension-eight SMEFT contributions on the Lam-Tung relation have been explored in~\cite{Alioli:2018ljm,Alioli:2020kez,Li:2024iyj}. To gain insight on how the light-quark dipole operators involving a $Z$ boson modify the predictions for the angular distributions, we integrate the angular distribution~(\ref{eq:angular}) over the azimuthal angle $\phi \in [0, 2 \pi]$. This yields
\bea \label{eq:int0angular} 
\frac{d\sigma}{dp_{T,ll} \hspace{0.25mm} dy_{ll} \hspace{0.25mm} dm_{ll}^2 \hspace{0.25mm} d\hspace{-0.35mm} \cos \theta} = \frac{3}{8} \frac{d\sigma}{dp_{T,ll} \hspace{0.25mm} dy_{ll} \hspace{0.25mm} dm_{ll}^2} \, \Bigg [ \left ( 1 +\frac{A_0}{2} \right ) \! \left ( 1 + a_0 \cos^2 \theta \right ) + A_4 \cos \theta \hspace{0.5mm} \Bigg ] \, .
\eea
Integrating~(\ref{eq:angular}) over the polar angle $\cos \theta \in [-1, 1]$ instead results in:
\beq \label{eq:int2angular} 
\begin{split}
\frac{d\sigma}{dp_{T,ll} \hspace{0.25mm} dy_{ll} \hspace{0.25mm} dm_{ll}^2 \hspace{0.25mm} d\phi} = \frac{1}{2 \pi} \frac{d\sigma}{dp_{T,ll} \hspace{0.25mm} dy_{ll} \hspace{0.25mm} dm_{ll}^2} \, \Bigg [ & \hspace{0.5mm} 1 + a_2 \cos 2 \phi + \frac{3 \pi}{16} A_3 \cos \phi \\[2mm] 
& + \frac{A_5}{2} \sin 2 \phi + \frac{3 \pi}{16} A_7 \sin \phi \hspace{0.25mm} \Bigg ] \, .
\end{split}
\eeq
Here, we have introduced the abbreviations 
\beq \label{eq:C0C2}
a_0 = \frac{2 - 3 A_0}{2 + \hspace{-0.75mm} A_0} \,, \qquad a_2 = \frac{A_2}{4} \,. 
\eeq
Note that $A_0 = 0$ necessarily implies $a_0 = 1$, and the reverse is also true.

The formulas~(\ref{eq:int0angular}) and~(\ref{eq:int2angular}) can provide some insight on how BSM contributions can modify the expected angular dependence.
Let us start by examining the case of the tree-level photon contribution to NC~DY~production at the~LHC for a single lepton flavor, i.e., the $q \bar{q} \to \gamma^\ast \to l^+ l^-$ process. In the limit of massless external fermions, the corresponding differential cross section is given by
\beq \label{eq:qqphotonllXS}
\frac{d\sigma}{d \hat \Omega} = \frac{\alpha^2 \hspace{0.25mm} Q_q^2}{12 \hspace{0.25mm} \hat s} \left ( 1 + \cos^2 \hat \theta \right ) \,,
\eeq
where $Q_u = 2/3$ and $Q_d = -1/3$ represent the electromagnetic charge factors of the up and down quark, respectively. Recalling that for a $2 \to 2$ process the CM and the CS frame are identical, a comparison of~(\ref{eq:qqphotonllXS}) with~(\ref{eq:int0angular}) gives 
\beq \label{eq:A0A2photon}
A_0 = 0 \,, \qquad A_2 = 0 \,.
\eeq 
The same reasoning can also be applied to the light-quark dipole contributions to $q \bar{q} \to Z \to l^+ l^-$ scattering. In this case, we find
\beq \label{eq:dipoleXS}
\frac{d\sigma}{d \hat \Omega} = \frac{\alpha \left ( 1- 4 s_w^2 + 8 s_w^4 \right ) v^2 \left | C_q \right |^2}{192 \hspace{0.125mm} \pi \hspace{0.25mm} s_w^4 \hspace{0.25mm} c_w^2 \hspace{0.25mm} \Lambda^4} \hspace{0.25mm} \frac{\hat s^2}{\left ( \hat s - M_Z^2 \right )^2} \left ( 1 - \cos^2 \hat \theta \right ) \,,
\eeq
when all external fermions are treated as massless, implying 
\beq \label{eq:A0A2dipole}
A_0 \neq 0 \,, \qquad A_2 = 0 \,.
\eeq
This shows that at the tree level, the light-quark dipole operators lead to changes in the prediction for the angular coefficient $A_0$, which already imply $A_0 \neq A_2$. It is important to note that for $p_{T,ll} = 0$, the azimuthal symmetry of the scattering process dictates that~$A_2 = 0$~\cite{Collins:1977iv}. Therefore, in both the SM and BSM scenarios, the condition $A_2 = 0$ must hold for these simple cases. Observe that the angular dependence in~(\ref{eq:qqphotonllXS}) and~(\ref{eq:dipoleXS}) resembles that of the denominator and numerator of~(\ref{eq:kappalarges}), respectively. We add that the polar angle dependence found in~(\ref{eq:qqphotonllXS}) and~(\ref{eq:dipoleXS}) can also be understood by analyzing the chiralities or helicities of the external fermions involved in the underlying scattering processes. In fact, the sign difference in the $\cos^2 \hat \theta$ term is due to the dipole interactions flipping the chiralities of the incoming quarks, whereas the photon couples exclusively to quarks and antiquarks with the same chirality.

We finally emphasize that the dipole operators~(\ref{eq:LSMEFT}) are the only dimension-six SMEFT operators that alter the interactions between the EW gauge bosons and light quarks in a manner that violates the Lam-Tung relation~(\ref{eq:lamtung}) at LO in QCD --- see Appendix~\ref{app:vectoroperator} for an example-based proof of this claim. This feature, along with the energy enhancement in~(\ref{eq:defchiq}), explains why the focus of this work is on the light-quark dipole operators~(\ref{eq:LSMEFT}).

\section{Data and MC predictions for $\bm{Z}$+jet production}
\label{sec:MCsetup}

To derive constraints on $C_q/\Lambda^2$, we utilize the measurement of the normalized $p_{T,ll}$ distribution of DY~lepton pairs reported by ATLAS~\cite{ATLAS:2019zci}. These measurements were performed using LHC data collected at a CM energy of $\sqrt{s} = 13 \, {\rm TeV}$, with an integrated luminosity of $36.1 \, \rm{fb}^{-1}$. Both dimuon and dielectron final states are analyzed within a dilepton invariant mass window of $66 \, {\rm GeV} < m_{ll} < 116 \, {\rm GeV}$. The results are presented within a fiducial phase space designed to be close to the experimental acceptance, defined by lepton transverse momenta $p_{T,l} > 27 \, {\rm GeV}$ and lepton pseudorapidities~$|\eta_l| < 2.5$. 

A key ingredient for the extraction of the constraints on $C_q/\Lambda^2$ is the SM prediction for the normalized $p_{T,ll}$ spectrum, which serves as a background for the BSM contributions. For this, we utilize the SM prediction published alongside the experimental results in~\cite{ATLAS:2019zci}. This SM prediction is based on an NNLO QCD prediction for the $Z$+jet process at~${\cal O} (\alpha_s^3)$ obtained with {\tt NNLOJET} \cite{Gehrmann-DeRidder:2015wbt,Gehrmann-DeRidder:2016jns}, further supplemented by NLO~EW effects~\cite{Denner:2011vu}. The reported SM prediction incorporates an estimate of the theoretical uncertainty based on the complete NNLO QCD calculation. As outlined in Appendix~\ref{app:ZJSM}, we independently evaluated the potential effects of uncertainties arising from scale variations,\footnote{We thank Alexander Huss for providing us with the SM predictions up to NNLO in QCD, obtained using {\tt NNLOJET}, as well as the breakdown of factorization and renormalization scale variations.} the choice of $\alpha_s(M_Z)$, and the input parton distribution functions (PDFs). These uncertainties were subsequently combined to determine the total theoretical uncertainties, which were found to be slightly larger than those reported by ATLAS. Consequently, the limits we derive on $C_q/\Lambda^2$ are more conservative than those that would be obtained by relying solely on the ATLAS evaluation of the normalized~$p_{T,ll}$ spectrum and their associated uncertainty estimates. The~light-quark dipole corrections to the normalized $p_{T,ll}$ spectrum in $Z$+jet~production are calculated by means of a {\tt FeynRules~2.0}~\cite{Alloul:2013bka} implementation of the Lagrangian~(\ref{eq:LSMEFT}) in the~{\tt UFO}~format~\cite{Degrande:2011ua}. The~corresponding $Z$+jet events were generated at LO in QCD using {\tt MadGraph5\_aMCNLO}~\cite{Alwall:2014hca}, followed by parton showering with {\tt PYTHIA~8.2}\cite{Sjostrand:2014zea}. The protons in the~LHC collisions were modeled using the {\tt NNPDF31\_nnlo\_as\_01180}~\cite{NNPDF:2017mvq} PDFs. To efficiently generate SMEFT events with high~$p_{T,ll}$, we applied a generation bias in the form $\left (p_{T,j}/{\rm TeV} \right )^2$, where $p_{T,j}$ is the $p_T$ of the jet recoiling against the $Z$~boson. This approach was employed to enhance the statistical precision of our BSM samples. The~BSM samples were subjected to the experimental cuts outlined at the beginning of this section. To ensure that our MC generation does not introduce bias into our results, we computed the high-$p_{T,ll}$ tail of the $p_{T,ll}$ spectrum for $Z$+jet production at LO~QCD within the SM using our setup. The results show agreement at the percent level with the ${\cal O} (\alpha_s)$ prediction obtained using {\tt NNLOJET}. Note that, since the interference term between the SM and BSM contributions vanishes in the limit of massless external fermions, the two predictions for the normalized~$p_{T,ll}$~distributions can be directly added to obtain the full SM plus BSM~results, with the BSM contributions calculated at ${\cal O} (\alpha_s)$.

As previously discussed, we will then use the derived constraints on $C_q/\Lambda^2$ to evaluate the potential impact that these BSM contributions may have on the prediction of the angular coefficients in the NC~DY process. These angular coefficients have been measured in the vicinity of the $Z$-boson mass peak in~\cite{ATLAS:2016rnf}. The data analyzed correspond to $20.3 \, {\rm fb}^{-1}$ of LHC collisions at a CM~energy of $\sqrt{s} = 8 \, {\rm TeV}$. In that analysis, the lepton pair is required to have an invariant mass within the window of $80 \, {\rm GeV} < m_{ll} < 100 \, {\rm GeV}$, while results are made available differential in $y_{ll}$ and $p_{T, ll}$ (up to $600 \, {\rm GeV}$). The~analysis~\cite{ATLAS:2016rnf} measures all eight angular coefficients~$A_i$ and finds deviations between the SM prediction $\big($calculated at ${\cal O} (\alpha_s^2)$$\big)$ and the measured values for the difference $A_0 - A_2$ in the tail of the $p_{T,ll}$ spectrum, suggesting that either higher-order QCD or EW corrections, or BSM physics, are necessary to accurately describe the data. Although less significant, a similar trend was also observed by CMS in~\cite{CMS:2015cyj}. A phenomenological analysis of these angular coefficients was subsequently performed in~\cite{Gauld:2017tww}, including the impact of QCD corrections up to ${\cal O}(\alpha_s^3)$, which relied on the calculations presented in~\cite{Gehrmann-DeRidder:2015wbt,Gehrmann-DeRidder:2016jns}. Those corrections reduced the quoted tension but still fail to accurately describe the data for the difference $A_0 - A_2$ at high $p_{T, ll}$.

To evaluate the impact of light-quark dipole operators on the angular coefficients~$A_i$, we calculated the relevant $q\bar q \to l^+ l^- + g$ tree-level matrix elements, as well as the crossed channels, using a combination of the {\tt FeynArts}~\cite{Hahn:2000kx} and {\tt FormCalc}~\cite{Hahn:2016ebn} packages. The~resulting analytic expressions were subsequently incorporated into a private code for NC~DY~production, developed in the context of~\cite{Gauld:2021zmq}. The BSM predictions were produced using {\tt PDF4LHC15\_nnlo\_30} PDFs, the same choice of EW inputs as in~\cite{Gauld:2017tww}, and were required to satisfy the same analysis criteria imposed by ATLAS in~\cite{ATLAS:2016rnf}. To determine the angular coefficients $A_i$, we employ projectors (see, for example,~\cite{Gauld:2017tww}). This~approach involves calculating normalized weighted averages over the angular variables $\theta$ and $\phi$. The~normalization is based on the LO QCD prediction of the $p_{T,ll}$ spectrum in $Z$+jet production. A~key challenge of this computation is that evaluating the required expectation values involves averaging over oscillatory functions of $\theta$ and $\phi$ --- for instance, the projector for~$A_2$ includes $\sin \theta$ and $\cos 2 \phi$ --- and that the calculation of the combination $A_0 - A_2$ suffers from numerical cancellations both in the SM and beyond. To address these challenges, the aforementioned private code was utilized, and an adaptive sampling for the integral was applied separately when computing $A_0$ and~$A_2$. This approach proved to be significantly more efficient than computing the angular coefficients $A_i$ using {\tt MadGraph5\_aMCNLO}, which was used to compute the normalized $p_{T,ll}$ spectra. Finally, note that the SM and BSM contributions to the angular coefficients $A_i$ are additive, as the corresponding amplitudes do not interfere when the external fermions are taken to be~massless. This allows the LO BSM predictions for the difference of the angular coefficients $A_0$ and $A_2$ to be directly added to the SM prediction, which is available at~${\cal O} (\alpha_s^3)$ from~\cite{Gauld:2017tww}.

\section{Numerical results}
\label{sec:numerics}

In this section, we derive the constraints on the Wilson coefficients of the light-quark dipole operators using the partial $Z$-boson decay widths and the normalized $p_{T,ll}$ spectrum in $Z$+jet production. With the derived limits, we then evaluate the maximum possible impact that this type of BSM effects could have on the angular coefficients relevant to the~Lam-Tung relation.

\subsection*{EW~precision measurements}
\label{sec:numericsZwidth}

The light-quark dipole interactions~(\ref{eq:LSMEFT}) can be constrained through the EW~precision measurements conducted at the $Z$-pole. In the following, we analyze the partial $Z$-boson decay widths. Without assuming lepton universality, the partial decay width of the $Z$~boson into light quarks has been measured with a precision of $5.9 \permil$ at the $95\%$~confidence~level~(CL)~\cite{ALEPH:2005ab,ParticleDataGroup:2024cfk}. This measurement, combined with the results in~(\ref{eq:Zwidths}), imposes the following condition:
\beq \label{eq:2DZwidth}
\frac{\left ( 0.46 \, {\rm TeV} \right )^2}{\Lambda^2} \hspace{0.5mm} \sqrt{\left | C_u \right |^2 + 0.93 \left | C_d \right |^2} < 7.6 \cdot 10^{-2} \, .
\eeq
Assuming only one Wilson coefficient is non-zero at a time, this inequality leads to the constraints:
\beq \label{eq:EWPOsbounds}
\frac{\left | C_u \right |}{\Lambda^2} < \frac{1}{\left ( 1.7 \, {\rm TeV} \right)^2} \,, \qquad \frac{\left | C_d \right |}{\Lambda^2} < \frac{1}{\left ( 1.6 \, {\rm TeV} \right )^2} \,. 
\eeq

\subsection*{Normalized $\bm{p_{T,ll}}$ distribution}
\label{sec:numericsptll}

\begin{figure}[t]
\begin{center}
\includegraphics[scale=0.85]{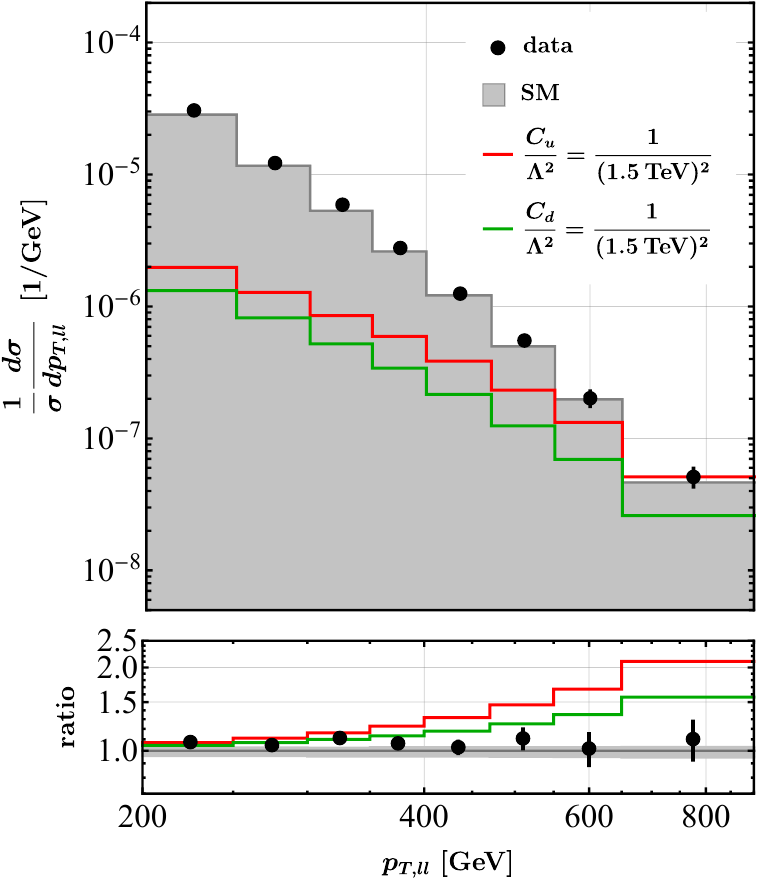}
\end{center}
\vspace{-2mm} 
\caption{\label{fig:ptllplot} Comparison of the normalized $p_{T,ll}$ distribution for $p_{T,ll} \in [200, 900] \, {\rm GeV}$. Black~points represent the ATLAS measurement~\cite{ATLAS:2019zci}, with statistical uncertainties depicted as black bars. The gray histogram corresponds to the SM prediction, and its systematic uncertainties are shown in the lower ratio plot as a gray band. Predictions for the BSM effects are displayed for $C_u/\Lambda^2 = 1/(1.5 \, {\rm TeV})^2$ and $C_d/\Lambda^2 = 1/(1.5 \, {\rm TeV})^2$, and depicted as red and green curves, respectively. The other Wilson coefficient not indicated is set to zero. Further details are provided in the main~text.}
\end{figure}

In~Figure~\ref{fig:ptllplot}, we compare the results of ATLAS~\cite{ATLAS:2019zci} with different predictions for the normalized~$p_{T,ll}$ spectrum in $Z$+jet production. The~uncertainties of the measurement are represented by black bars. The SM prediction is represented by a gray histogram, with its uncertainties illustrated as a gray band in the lower ratio plot. These predictions are taken from~\cite{ATLAS:2019zci}, where they were calculated using {\tt NNLOJET}~\cite{Gehrmann-DeRidder:2015wbt,Gehrmann-DeRidder:2016jns} and supplemented with NLO EW corrections~\cite{Denner:2011vu}. The red and green curves represent our BSM predictions for the normalized~$p_{T,ll}$ distribution for the two choices $C_u/\Lambda^2= 1/(1.5 \, {\rm TeV})^2$ and $C_d/\Lambda^2 = 1/(1.5 \, {\rm TeV})^2$ of Wilson coefficients introduced in~(\ref{eq:killphoton}). These results were generated using the MC~setup detailed in~Section~\ref{sec:MCsetup}. In line with~(\ref{eq:kappalarges}), we observe that the considered BSM effects become more pronounced relative to the SM background at high~$p_{T,ll}$. For example, in the highest bin, $p_{T,ll} \in [650, 900] \, {\rm GeV}$, the benchmark values of~$C_q/\Lambda^2$ result in enhancements of approximately~$210\%$ and~$160\%$ relative to the SM prediction. It is evident from the figure that such significant enhancements are ruled out by the ATLAS~data.

We proceed to derive $95\%$ CL limits on $C_q/\Lambda^2$, using the results for the normalized~$p_{T,ll}$ spectrum in $Z$+jet production presented in~Figure~\ref{fig:ptllplot}. The significance is determined as a ratio of Poisson likelihoods, adjusted to account for the statistical and systematic uncertainties on the background reported in~\cite{ATLAS:2019zci} and a systematic uncertainty of~$5\%$ on the BSM predictions. In Appendix~\ref{app:ZJSM}, we assess the systematic uncertainty of the SM prediction. Our evaluation yields slightly larger uncertainties than those reported in~\cite{ATLAS:2019zci}. As~a~result, the limits we derive on the combinations $C_q/\Lambda^2$ are more conservative than those that would be obtained using only the ATLAS evaluation of the normalized $p_{T,ll}$ spectrum and its associated uncertainty estimates. All systematic uncertainties are incorporated as Gaussian constraints~\cite{Cowan:2010js}. Our likelihood analysis leads to the following inequality:
\beq \label{eq:2DZplusjet}
\frac{\left ( 0.46 \, {\rm TeV} \right)^2}{\Lambda^2} \hspace{0.5mm} \sqrt{\left | C_u \right |^2 + 0.51 \left | C_d \right |^2} < 4.1 \cdot 10^{-2} \, .
\eeq
Observe that we have expressed~(\ref{eq:2DZplusjet}) in a form similar to~(\ref{eq:2DZwidth}) to illustrate that current LHC measurements of the high-$p_{T,ll}$ spectrum in DY production exhibit a sensitivity to $C_q/\Lambda^2$ that is already stronger by a factor of approximately 1.8 (1.4) compared to the SLC and LEP measurements of the partial decay width of the $Z$ boson into up quarks~(down quarks). Under~the~assumption that only one Wilson coefficient is non-zero, the condition~(\ref{eq:2DZplusjet}) translates into the following upper~limits:
\beq \label{eq:pTllbounds}
\begin{split}
\frac{|C_u|}{\Lambda^2} < \frac{1}{\left ( 2.3 \, {\rm TeV} \right)^2} \,, \qquad \frac{|C_d|}{\Lambda^2} < \frac{1}{\left ( 1.9 \, {\rm TeV} \right)^2} \,.
\end{split}
\eeq
A few comments appear to be appropriate. First, since the SM predictions in~Figure~\ref{fig:ptllplot} consistently fall below the data, the limits~(\ref{eq:pTllbounds}) depend on how many $p_{T,ll}$ bins are included in the likelihood analysis. The values reported above are based on the two highest $p_{T,ll}$ bins from the ATLAS measurements~\cite{ATLAS:2019zci}, as this choice (see Appendix~\ref{app:fitdetails}) yields the most stringent upper limits on $|C_q|/\Lambda^2$. We observe that our likelihood analysis also results in lower bounds with $|C_q|/\Lambda^2 > 0$, which challenge the SM. However, these bounds arise from the SM prediction for the normalized~$p_{T,ll}$ spectrum consistently falling below the ATLAS data by an almost constant offset, whereas the considered BSM effects would scale as~$p_{T,ll}^2$. Thus,~we restricted our analysis to determining only upper limits on~$|C_q|/\Lambda^2$, following~\cite{Cowan:2010js}, since these are less affected by the mismatch between the SM prediction and the data than the lower bounds.

Second, the limits in~(\ref{eq:pTllbounds}) are stronger than those in~(\ref{eq:EWPOsbounds}), a result consistent with the findings of~\cite{daSilvaAlmeida:2019cbr}. The bounds in~(\ref{eq:pTllbounds}) are primarily constrained by the limited statistics in the current ATLAS data on $Z$+jet production at high $p_{T,ll}$. Since~this measurement uses only $36.1 \, {\rm fb}^{-1}$ of data, naive luminosity scaling suggests that at the high-luminosity option of the LHC~(HL-LHC), with $3000 \, {\rm fb}^{-1}$ of integrated luminosity, an improvement by a factor of approximately 3 can be expected concerning~(\ref{eq:pTllbounds}). To make this statement more precise, we repeated our likelihood analysis using simulated SM and BSM samples, assuming a CM energy of $\sqrt{s} = 14 \, {\rm TeV}$ and considering $p_{T,ll}$ values in the range $[600, 3000] \, {\rm GeV}$. We assumed a $5\%$ systematic uncertainty for both the SM and the BSM predictions. Our analysis indicates that the bounds in~(\ref{eq:pTllbounds}) could potentially be improved by a factor of about $4.5$ at the HL-LHC. The improvement exceeds naive expectations due to the quadratic energy growth of the light-quark dipole corrections to the $p_{T,ll}$ distribution~$\big($see~(\ref{eq:kappalarges})$\big)$, which offsets the suppression from the decreasing parton~luminosities.

\subsection*{Violation of Lam-Tung relation}
\label{sec:numericslumtungl}

\begin{figure}[t]
\begin{center}
\includegraphics[scale=0.85]{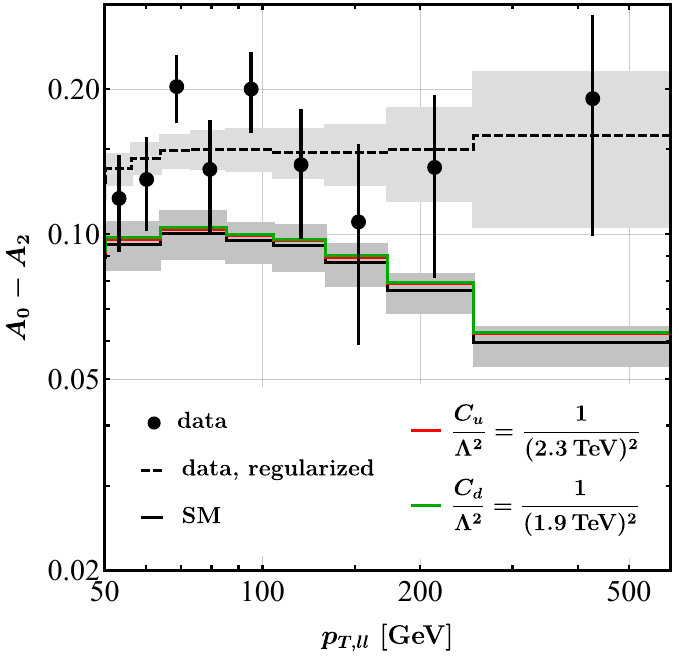}
\end{center}
\vspace{-2mm} 
\caption{\label{fig:lamtungplot} Comparison of the predictions for $A_0 - A_2$ in the range $p_{T,ll} \in [50, 600] \, {\rm GeV}$. The~black points show the central values of the ATLAS measurement~\cite{ATLAS:2016rnf}, with statistical uncertainties represented by black error bars. The black solid curve and gray band represent the SM prediction and its systematic uncertainties. Predictions including BSM effects are shown for $C_u/\Lambda^2 = 1/(2.3 \, {\rm TeV})^2$ and $C_d/\Lambda^2 = 1/(1.9 \, {\rm TeV})^2$, depicted as red and green curves, respectively. The other Wilson coefficient not specified in the figure is set to zero. The regularized ATLAS data is also included and shown as a dashed black line and a light gray band. Additional details are given in the main~text.}
\end{figure}

Using the limits in~(\ref{eq:EWPOsbounds}) and~(\ref{eq:pTllbounds}), we can now evaluate the maximum effect of light-quark dipole operators on the prediction of the difference between the angular coefficients $A_0$ and $A_2$, and thus assess their potential to explain the observed violation of the Lam-Tung relation, which exceeds the NNLO QCD prediction within the SM. In Figure~\ref{fig:lamtungplot}, we show our results for the combination $A_0 - A_2$ of angular coefficients as a function of $p_{T,ll}$ for $pp$~collisions at a CM~energy of $\sqrt{s} = 8 \, {\rm TeV}$. The~ATLAS data~\cite{ATLAS:2016rnf} is shown as black points with error bars. The~SM~prediction is depicted by a solid black line, with its uncertainties represented by a gray band. As detailed in Section~\ref{sec:MCsetup}, these predictions are based on the NNLO~QCD calculation performed in~\cite{Gauld:2017tww}. The plot reveals a tendency for the ATLAS data to systematically exceed the SM prediction for $A_0 - A_2$ at higher $p_{T,ll}$ values. A~$\chi^2$~test performed in~\cite{Gauld:2017tww}, however, demonstrated that the SM prediction aligns reasonably well with the ATLAS measurement across all 38 data points, yielding $\chi^2/38 = 1.8$. For~a better comparison with the recent article~\cite{Li:2024iyj}, we have included in~Figure~\ref{fig:lamtungplot} also the regularized ATLAS data, shown as a dashed black line with gray shading --- details on the experimental regularization procedure can be found in~Appendix~C~of~\cite{ATLAS:2016rnf}. For the purposes of this work, it is sufficient to note that the regularization procedure introduces large bin-to-bin correlations in the distributions of the angular coefficients $A_i$. A visual comparison with the regularized data can therefore be misleading, as these strong correlations are not apparent in the regularized histograms. For~$A_0 - A_2$, the figure clearly illustrates that a naive comparison of the SM prediction with the regularized ATLAS data will lead to an overestimation of the disagreement between theory and experiment.

The red and green curves in~Figure~\ref{fig:lamtungplot} illustrate our $A_0 - A_2$ predictions for the two choices of Wilson coefficients, $C_u/\Lambda^2 = 1/(2.3 \, {\rm TeV})^2$ and $C_d/\Lambda^2 = 1/(1.9 \, {\rm TeV})^2$, corresponding to the upper limits given in~(\ref{eq:pTllbounds}). Note that the value of the Wilson coefficient for the up-quark dipole operator considered in the paper~\cite{Li:2024iyj} corresponds to $C_u/\Lambda^2 = s_w/{\rm TeV}^2 \simeq 1/(1.4 \, {\rm TeV})^2$. Since the BSM contributions to $A_0 - A_2$ scale as~$(|C_q|/\Lambda^2)^2$, the choice of Wilson coefficient for the up-quark dipole operator in~\cite{Li:2024iyj} results in relative effects approximately six times larger than those displayed in~Figure~\ref{fig:lamtungplot}. This feature is illustrated in Appendix~\ref{app:more}. The~figure shows that the inclusion of the considered light-quark dipole contributions has only a very minor impact on the $A_0 - A_2$ distribution, staying within the uncertainty band of the SM prediction. Consequently, the resulting BSM effects are clearly insufficient to bridge the gap between theory and experiment in the final bin, spanning the range~$[253, 600] \, {\rm GeV}$, of the ATLAS measurement. The $C_q/\Lambda^2$ values shown in the plot therefore nicely highlight that, given current experimental constraints, BSM effects as described by~(\ref{eq:LSMEFT}) are essentially ruled out as a solution to the tension between the SM prediction and the ATLAS measurement at large~$p_{T,ll}$.

\section{Conclusions}
\label{sec:conclusions}

In this article, we conducted model-independent analyses of potential BSM modifications in NC~DY production in both $e^+ e^-$ and $pp$ collisions, focusing on light-quark dipole operators arising at the dimension-six level within the SMEFT framework. These analyses allowed us to revisit the findings of~\cite{daSilvaAlmeida:2019cbr,Li:2024iyj}. Previous studies have derived constraints on light-quark dipole couplings using EW~precision measurements at the $Z$-pole and analyses of EW diboson and DY production at the LHC~\cite{daSilvaAlmeida:2019cbr}. Additionally, this type of BSM physics has been proposed as a possible explanation for the observed discrepancies between theoretical predictions and experimental measurements of the violation of the Lam-Tung~relation in the tail of the $p_{T,ll}$ spectrum in $Z$+jet production~\cite{Li:2024iyj}.

We began our numerical analysis by deriving constraints on the Wilson coefficients of the light-quark dipole operators using precision measurements of the $Z$-boson decay width performed at SLC~and~LEP, as well as the normalized $p_{T,ll}$ distribution measured in $pp \to \gamma^\ast/Z + X \to l^+ l^- + X$ production at the LHC. The latter bounds were computed using existing state-of-the-art predictions for NC~DY production within the SM, incorporating NNLO~QCD corrections~\cite{Gehrmann-DeRidder:2015wbt,Gehrmann-DeRidder:2016jns} and NLO EW effects~\cite{Denner:2011vu}, alongside a conservative estimate of theoretical uncertainties --- details of which can be found in~Appendix~\ref{app:ZJSM}. We~observed that the effects of light-quark dipole operators are enhanced at high energies, and as a result, the precision of current $p_{T,ll}$ measurements at the LHC already exceeds the sensitivity of the high-precision $e^+ e^- \to Z \to q \bar q$ measurements at SLC and LEP. This~conclusion is in agreement with~\cite{daSilvaAlmeida:2019cbr}.

Next, we applied the obtained constraints to evaluate the maximum impact that light-quark dipole operators could have on the predictions for the angular coefficients $A_0$ and~$A_2$, which appear in the Lam-Tung relation~(\ref{eq:lamtung}). Our BSM predictions for the difference $A_0 - A_2$ were derived at LO and combined with the existing SM predictions for the angular coefficients $A_i$ in $Z$-boson production at NNLO in QCD~\cite{Gehrmann-DeRidder:2015wbt,Gehrmann-DeRidder:2016jns,Gauld:2017tww}, which represent the most advanced SM calculations currently available. We found that the limits~(\ref{eq:EWPOsbounds}) and~(\ref{eq:pTllbounds}) on the Wilson coefficients of the light-quark dipole operators exclude the parameter space that could account for the discrepancy between theoretical predictions and experimental observations of the Lam-Tung relation, as reported by the ATLAS collaboration in~\cite{ATLAS:2016rnf}. This finding contrasts with the results of the recent work~\cite{Li:2024iyj}, which relied on values for the Wilson coefficients that are inconsistent with both~(\ref{eq:EWPOsbounds}) and~(\ref{eq:pTllbounds}). Therefore, BSM effects of the type~(\ref{eq:LSMEFT}) are essentially ruled out as the cause of the excess observed in the difference $A_0 - A_2$ at high $p_{T,ll}$. This suggests that other factors, such as unaccounted-for QCD, EW, or experimental effects, are more likely explanations at present.

We also noted that the constraints in~(\ref{eq:pTllbounds}) on the Wilson coefficients of the light-quark dipole operators, derived from current LHC data on the normalized $p_{T,ll}$ spectrum in $Z$+jet production, are limited by statistics. This is because the bounds are primarily driven by the high $p_{T,ll}$ bins, where the BSM contributions have a larger impact compared to the SM. As a result, future measurements of NC~DY production during the HL-LHC could potentially improve the constraints presented in~(\ref{eq:pTllbounds}) by a factor of about~4.5. Similar statements apply to the HL-LHC measurements of all angular coefficients~$A_i$, which are currently not well measured in the high-$p_{T,ll}$ regime due to insufficient statistics. Precise future experimental measurements of the Lam-Tung relation could therefore offer additional and partially complementary insights into BSM physics, especially if deviations are detected at high~$p_{T,ll}$ in the unpolarized differential cross section.

\section*{Acknowledgements} We thank Alexander Huss for providing us with both the SM predictions for the differential $p_{T,ll}$ spectrum and the fiducial cross section in NC~DY production, obtained using {\tt NNLOJET}, including the breakdown of factorization and renormalization scale variations. We also acknowledge Luc~Schnell's support in event generation with {\tt MadGraph5\_aMCNLO}. The~Feynman diagrams shown in this work were generated and drawn with {\tt FeynArts}. JW is part of the International Max Planck Research School (IMPRS) on “Elementary Particle Physics”. 

\begin{appendix}
\renewcommand{\theequation}{\thesection.\arabic{equation}}
\setcounter{equation}{0}

\section{SM prediction for $\bm{Z}$+jet production}
\label{app:ZJSM}

In this appendix, we estimate the theoretical uncertainties affecting the $p_{T,ll}$ spectrum in $Z$+jet production within the SM. These uncertainties arise from various sources, including variations in the renormalization and factorization scales, uncertainties in the value of the strong coupling constant $\alpha_s(M_Z)$ at the scale $M_Z$, and uncertainties associated with the chosen PDF set. Each of these sources of theoretical uncertainty will be discussed individually before determining the total theoretical uncertainties in the $p_{T,ll}$ distribution.

\subsection*{Scale uncertainties}
\label{app:uncscale}

The scale uncertainties of the $p_{T,ll}$ spectrum in the SM are evaluated using the $Z$+jet production results at ${\cal O} (\alpha_s^3)$, as implemented in {\tt NNLOJET}~\cite{Gehrmann-DeRidder:2015wbt,Gehrmann-DeRidder:2016jns}. One of the authors of these studies provided the {\tt NNLOJET} predictions for the $p_{T,ll}$ spectrum and fiducial cross section, including variations in factorization and renormalization scales. Similar results were also shared with ATLAS for the measurement reported in~\cite{ATLAS:2019zci}. Since these results play a crucial role in determining the limits presented in Section~\ref{sec:numerics}, we summarize their key details below. The predictions were derived using the central member of the {\tt NNPDF31\_nnlo\_as\_01180} PDF set. In this calculation, the renormalization and factorization scales were dynamically set on an event-by-event basis to
\beq\label{eq:scalechoices}
\mu_R = k \hspace{0.5mm} E_{T,Z} \, , \qquad \mu_F = m \hspace{0.5mm} E_{T,Z} \, ,
\eeq
where $E_{T,Z} = \sqrt{m_{ll}^2 + p_{T,ll}^2}$ represents the transverse energy of the virtual $Z$ boson. Seven combinations of $k$ and $m$ are considered:
\beq \label{eq:kmsets}
\{k,m\} \in \Big \{ \{1,1\} , \{0.5,0.5\} , \{2,2\} , \{1,0.5\} , \{1,2\} , \{0.5,1\} , \{2,1\} \Big \} \, .
\eeq
The scale choice $\{k, m\} = \{1, 1\}$ is used to define the central values of the predictions, including the total cross section and the $p_{T,ll}$ distribution. The theoretical uncertainties due to scale variations are determined by taking the envelope of the predictions corresponding to all seven combinations in~(\ref{eq:kmsets}). For instance, in the case of the total $Z$-boson production cross section at ${\cal O} (\alpha_s^2)$ at a CM energy of $\sqrt{s} = 13 \, {\rm TeV}$, within the fiducial region defined at the beginning of Section~\ref{sec:MCsetup}, the following result for a single lepton flavor is obtained:
\beq \label{eq:sigmascale}
\sigma = 728.7 \left (1 ^{+0.42\%}_{-0.72\%} \right ) {\rm pb} \,.
\eeq
The same approach, when applied to the case of the $p_{T,ll}$ spectrum, yields relative variations of about $^{+2\%}_{-3\%}$ within the $p_{T,ll}$ range of interest. This feature is illustrated in~Figure~\ref{fig:allunc} by the solid red~lines.

\subsection*{Uncertainties related to $\bm{\alpha_s}$}
\label{app:uncalpha}

The uncertainties associated with the choice of the strong coupling constant $\alpha_s$ defined at the scale $M_Z$ are estimated by generating $Z$+jet event samples at LO in QCD using {\tt MadGraph5\_aMCNLO} with various PDF fits from {\tt NNPDF31\_nnlo}, each obtained with a different value of $\alpha_s(M_Z)$. The following eleven choices are considered:
\bea \label{eq:alphaschoices}
\alpha_s (M_Z) \in \big \{ 0.108, 0.110, 0.112, 0.114, 0.116, 0.117, 0.118, 0.119, 0.120, 0.122, 0.124 \big \} \, . \hspace{2mm} 
\eea
The uncertainties of the predictions are derived by applying a polynomial fit to the results, considering the variations in the strong coupling constant $\alpha_s(M_Z) = 0.1180 \pm 0.0009$~\cite{ParticleDataGroup:2024cfk}. The central values of the predictions are identified with the values obtained for the central member of the {\tt NNPDF31\_nnlo\_as\_01180} PDF set. Using the fiducial cross section of NC~DY production at a CM energy of $\sqrt{s} = 13 \, {\rm TeV}$ as an example, we find that this method yields the following prediction for a single lepton flavor:
\beq \label{eq:sigmaalphas}
\sigma = 710.1 \left (1 ^{+0.37\%}_{-0.65\%} \right ) {\rm pb} \,.
\eeq
As depicted by the dotted green curves in~Figure~\ref{fig:allunc}, this approach results in relative uncertainties of less than about $\pm 0.5\%$ for the $p_{T,ll}$ distribution, which are associated with the choice of the strong coupling constant $\alpha_s$.

\subsection*{PDF uncertainties}
\label{app:uncpdf}

The PDF uncertainties are determined by generating $Z$+jet predictions at LO in QCD using {\tt MadGraph5\_aMCNLO} for all 100 members of the {\tt NNPDF31\_nnlo\_as\_01180} PDF set. Assuming a Gaussian distribution, the central values of our predictions are determined from the results corresponding to the central member of the {\tt NNPDF31\_nnlo\_as\_01180} PDF set, while the uncertainties are calculated from the standard deviations relative to the average of all 100 members. Applying this methodology to the fiducial cross section for NC~DY~production at a CM energy of $\sqrt{s} = 13 \, {\rm TeV}$ gives
\beq \label{eq:sigmaPDF}
\sigma = 710.1 \left (1 ^{+0.85\%}_{-0.74\%} \right ) {\rm pb} \,,
\eeq
for a single lepton flavor. In the case of the $p_{T,ll}$ spectrum, the corresponding PDF uncertainties are depicted as dashed blue lines in~Figure~\ref{fig:allunc}. We find that the relative uncertainties due to the choice of PDFs are approximately $\pm 1\%$ at low $p_{T,ll}$ values, rising to around~$\pm 1.5\%$ in the high$\hspace{0.25mm}$-$\hspace{0.25mm} p_{T,ll}$ tail of the shown $Z$+jet results.

\subsection*{Combined theoretical uncertainties}
\label{app:unctot}

\begin{figure}[t]
\begin{center}
\includegraphics[scale=0.85]{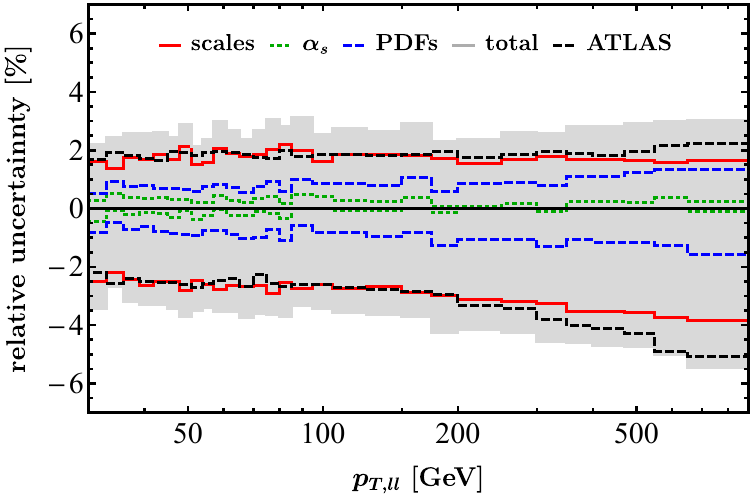}
\end{center}
\vspace{-2mm} 
\caption{\label{fig:allunc} The individual uncertainties and their combined effect for the $p_{T,ll}$ spectrum in $Z$+jet production at the LHC with a CM energy of $\sqrt{s} = 13 \, {\rm TeV}$. The uncertainties due to scale variations, $\alpha_s$, and PDFs are illustrated by the solid red, dotted green, and dashed blue lines, respectively, while the total combined uncertainties are represented by the gray band. For comparison, the total uncertainties reported by the ATLAS measurement~\cite{ATLAS:2019zci} are shown as dashed black lines.}
\end{figure}

We calculate the total theoretical uncertainties by combining the uncertainties from $\alpha_s$ and the PDFs in quadrature, while adding the uncertainties from scale variations linearly. Note that combining the uncertainties from $\alpha_s$ and the PDFs in quadrature aligns with the recommendations for using PDF sets as formulated in~\cite{Butterworth:2015oua}. In fact, assuming Gaussian distributions and linear error propagation, adding the $\alpha_s$ and PDF uncertainties in quadrature automatically accounts for the correlation between these two sources of uncertainties~\cite{Lai:2010nw}, making this approach well justified. In contrast, adding the scale uncertainties linearly to the combined $\alpha_s$ and PDF uncertainties is simply one option. This approach is justified by recognizing that scale uncertainties are systematic and non-Gaussian in~nature. 

Figure~\ref{fig:allunc} presents the individual uncertainties and their combination for the $p_{T,ll}$ spectrum in $Z$+jet~production, assuming LHC collisions at a CM energy of $\sqrt{s} = 13 \, {\rm TeV}$. We~observe that scale variations dominate the uncertainties across all bins, with PDF uncertainties becoming slightly more important at higher $p_{T,ll}$ values. As a result, the total theoretical uncertainties grow from $^{+2.2\%}_{-3.4\%}$ at $p_{T,ll} = 30 \, {\rm GeV}$ to $^{+3.0\%}_{-5.4\%}$ at $p_{T,ll} = 600 \, {\rm GeV}$. The ATLAS analysis~\cite{ATLAS:2019zci} instead quotes total theoretical uncertainties of $^{+1.7\%}_{-2.2\%}$ and $^{+2.2\%}_{-5.1\%}$ for the same $p_{T,ll}$ values --- the total uncertainties from the ATLAS measurement~\cite{ATLAS:2019zci} are displayed as dashed black lines in~Figure~\ref{fig:allunc}. These numbers imply that, when symmetrized, our uncertainty estimates are approximately $20\%$ larger than those reported by ATLAS. The constraints we established in~Section~\ref{sec:numerics} on $|C_q|/\Lambda^2$ therefore turn out to be more conservative than those that would derive from the ATLAS evaluation of the theoretical uncertainties affecting the normalized~$p_{T,ll}$ spectrum. We note that the theoretical uncertainties associated with the EW corrections to $Z$+jet production are not accounted for in our uncertainty estimate. These uncertainties are at the level of $1\%$ at $p_{T,ll} = 1 \, {\rm TeV}$~\cite{Lindert:2017olm}, and therefore would have an insignificant impact on our likelihood analysis used to derive the limits on the combinations $C_q/\Lambda^2$.

\section{Details on fit to $\bm{Z}$+jet data}
\label{app:fitdetails}

\begin{figure}[t]
\begin{center}
\includegraphics[scale=0.85]{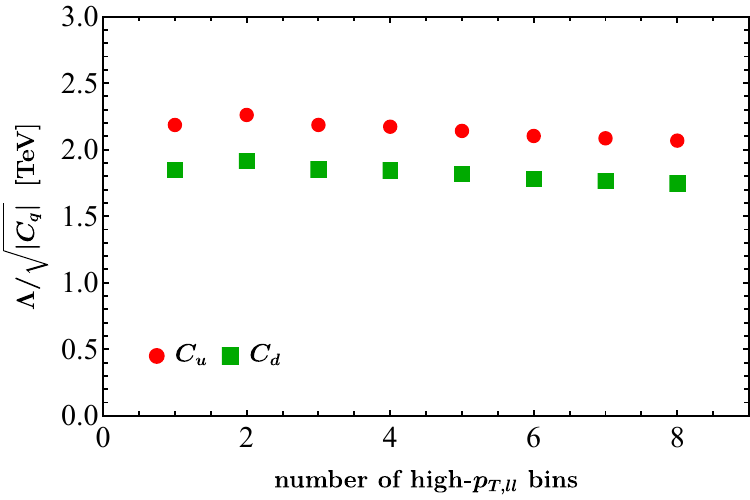}
\end{center}
\vspace{-2mm} 
\caption{\label{fig:fitplot} $95\%$ CL lower limits on $\Lambda/\sqrt{|C_q|}$ are shown as a function of the number of high-$p_{T,ll}$ bins from the ATLAS measurement~\cite{ATLAS:2019zci} included in the statistical analysis outlined in~Section~\ref{sec:numerics}. The results for $C_u$ and $C_d$ are represented by red dots and green squares, respectively. Additional details are provided in the main text.}
\end{figure}

In this appendix, we present details on the Poisson likelihood analysis that has been used in~Section~\ref{sec:numerics} to derive limits on~$C_q/\Lambda^2$ using the ATLAS data~\cite{ATLAS:2019zci} on $Z$+jet production. Figure~\ref{fig:fitplot} illustrates the $95\%$ CL lower limits on~$\Lambda/\sqrt{|C_q|}$ as a function of the number of~high-$p_{T,ll}$ bins from the ATLAS measurement included in our statistical analysis. The~results for $C_u$ and $C_d$ are depicted by red dots and green squares, respectively. The~first observation is that the derived lower bounds on $\Lambda/\sqrt{|C_q|}$ show little dependence on the number of high-$p_{T,ll}$ bins included in our Poisson likelihood analysis. The dependence of the limits on the number of bins can be qualitatively explained by noting that when only the highest bin is included in the fit, the limited statistical precision of the ATLAS measurement largely determines the resulting bound. Adding more high-$p_{T,ll}$ bins generally enhances the bounds on $\Lambda/\sqrt{|C_q|}$. However, the improvement is limited because the SM prediction remains consistently below the ATLAS data throughout the entire $p_{T,ll}$ range of $[200, 900] \, {\rm GeV}$, as illustrated in Figure~\ref{fig:ptllplot}. The values presented in~(\ref{eq:pTllbounds}) are derived from the two highest $p_{T,ll}$ bins, as this choice provides the most stringent upper limits on~$|C_q|/\Lambda^2$. It is worth pointing out that our likelihood analysis also leads to lower bounds where~$|C_q|/\Lambda^2 > 0$. However, these limits stem from the fact that the SM prediction for the normalized $p_{T,ll}$ spectrum is consistently lower than the ATLAS data, potentially due to unaccounted-for QCD, EW, or experimental effects, which could bias the results of the likelihood analysis. Therefore, we limited our fit to establishing upper limits on~$|C_q|/\Lambda^2$, as these are less influenced by the discrepancy between theory and experiment compared to the lower~bounds.

\section{On other SMEFT contributions to $\bm{Z}$+jet production}
\label{app:vectoroperator}

In this appendix, we present a brief discussion of another type of SMEFT contributions to $Z$+jet production. For a more in-depth analysis, including contributions from both dimension-six and dimension-eight operators, see for instance~\cite{Alioli:2018ljm,Alioli:2020kez}. As an explicit example of a dimension-six operator that modifies the couplings between the $Z$ boson and light quarks, we consider the following effective interaction:
\beq \label{eq:vector}
\mathcal{L}_{\rm SMEFT} \supset \frac{C_{Hu}}{\Lambda^2} \hspace{0.5mm} (H^\dagger i \overset{\leftrightarrow}{D}_\mu H) (\bar u \gamma^\mu u) \,.
\eeq 
In this expression, $H^\dagger i \overset{\leftrightarrow}{D}_\mu H = i H^\dagger \big (D_\mu - \overset{\leftarrow}{D}_\mu \big ) H$, where $D_\mu$ represents the standard covariant derivative. Additionally, we assume that the Wilson coefficient $C_{Hu}$ is real. 

To gain a qualitative insight into the effect of the interaction~(\ref{eq:vector}) on NC~DY production predictions, we evaluate the ratio of matrix elements defined in~(\ref{eq:defchiq}). We find the following simple analytic result: 
\beq \label{eq:chiuvector}
\chi_u = 1 - \frac{9 \hspace{0.25mm} v^2}{9 - 24 s_w^2 + 32 s_w^4} \left ( \frac{8 \hspace{0.25mm} s_w^2 \hspace{0.25mm} C_{Hu}}{3 \hspace{0.25mm} \Lambda^2} - \frac{v^2 \left | C_{Hu} \right |^2}{\Lambda^4} \right ) \,.
\eeq
Note that, in contrast to~(\ref{eq:chiq}), the ratio~(\ref{eq:chiuvector}) does not depend on the kinematical variables~$s$ and $t$. This observation has two implications. First, the on-shell production of a $Z$~boson and a jet resulting from~(\ref{eq:vector}) is not enhanced at high energies relative to the SM~background. Second, the effective interactions considered do not cause a violation of the Lam-Tung relation~(\ref{eq:lamtung}) at LO in QCD. 

\begin{figure}[t]
\begin{center}
\includegraphics[scale=0.575]{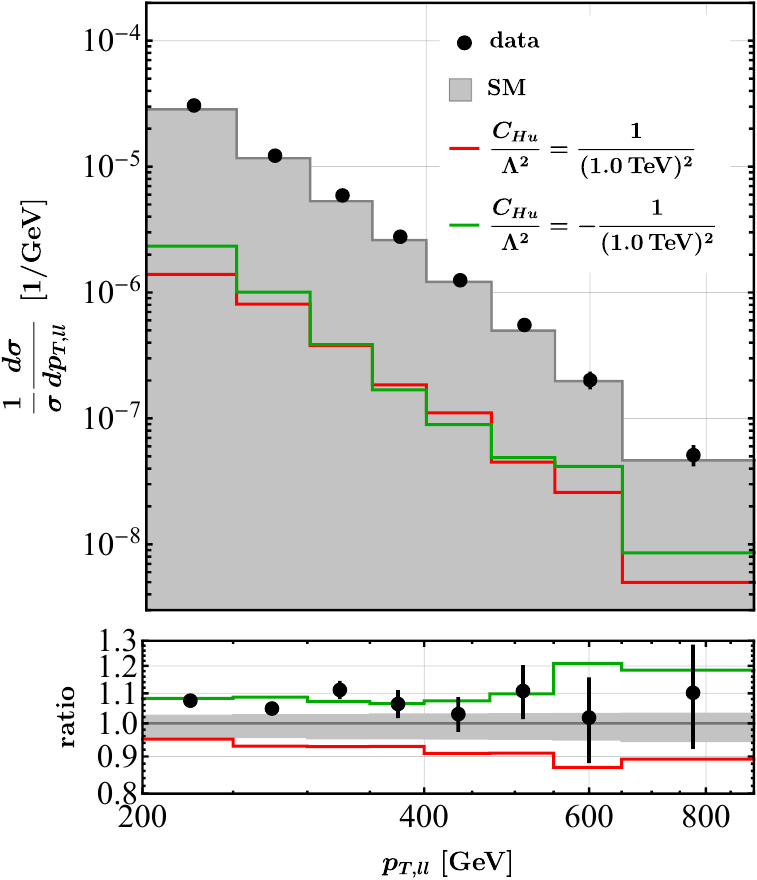} \qquad
\raisebox{10mm}[0mm][0mm]{\includegraphics[scale=0.575]{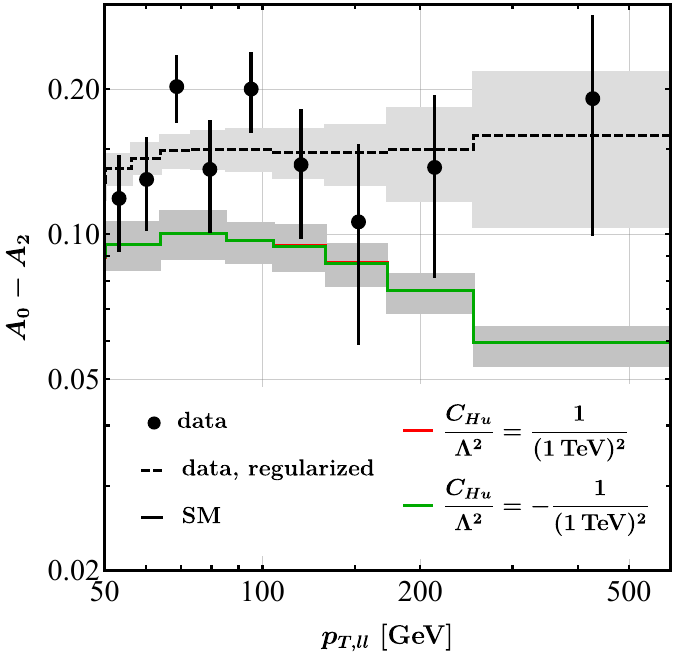}}
\end{center}
\vspace{-2mm} 
\caption{\label{fig:strom} Left: Similar to~Figure~\ref{fig:ptllplot}. Right: Similar to Figure~\ref{fig:lamtungplot}. Both plots employ the choices $C_{Hu}/\Lambda^2 = 1/(1 \, {\rm TeV})^2$ (solid red lines) and $C_{Hu}/\Lambda^2 = -1/(1 \, {\rm TeV})^2$ (solid green lines) of the Wilson coefficient appearing in~(\ref{eq:vector}). For additional explanations see the main text.}
\end{figure}

To support these statements, Figure~\ref{fig:strom} presents $Z$+jet predictions for two different values of the Wilson coefficient $C_{Hu}$. The left plot displays the BSM $p_{T,ll}$ spectra, generated using the MC setup described in~Section~\ref{sec:MCsetup}, shown as colored lines in comparison to the SM~background. The upper panel of the plot illustrates the magnitude of the BSM contribution to the distribution, while the lower panel presents the relative correction compared to the SM prediction. We present the magnitudes of the differential cross sections since, for $C_{Hu}/\Lambda^2 = 1/(1 \, {\rm TeV})^2$, the BSM contribution to the $p_{T,ll}$ spectrum is negative. Conversely, for $C_{Hu}/\Lambda^2 = -1/(1 \, {\rm TeV})^2$, the corrections are positive. Importantly, for both values of the Wilson coefficient introduced in~(\ref{eq:vector}), the BSM corrections remain nearly flat, with only a slight increase in magnitude as $p_{T,ll}$ increases. This behavior aligns nicely with~(\ref{eq:chiuvector}). Note that the chosen values for $C_{Hu}$ are viable in the SMEFT framework when no flavor assumptions are imposed~\cite{Gauld:2023gtb}. The right plot in Figure~\ref{fig:strom} instead displays the BSM predictions for $A_0 - A_2$ assuming again $C_{Hu}/\Lambda^2 = 1/(1 \, {\rm TeV})^2$ and $C_{Hu}/\Lambda^2 = -1/(1 \, {\rm TeV})^2$. The results shown demonstrate that the effective operator~(\ref{eq:vector}) does not cause any modifications to the Lam-Tung relation, within the statistical uncertainties, even for large Wilson coefficients. This feature is evident from~(\ref{eq:chiuvector}), which shows that~(\ref{eq:vector}) alters only the overall coupling strength between the $Z$ boson and right-handed up-quarks, without affecting the kinematics. As in the SM, BSM models that include effective interactions of the form~(\ref{eq:vector}) satisfy the relation (\ref{eq:lamtung}) beyond LO in QCD.

The above discussion has concentrated on the specific effective interactions~(\ref{eq:vector}), but it similarly applies to any other dimension-six operators that alter the couplings between the $Z$ boson and light quarks. The deviation patterns observed in~Figure~\ref{fig:strom} are therefore representative of BSM modifications of this type, and notably distinct from the SMEFT results displayed in~Figures~\ref{fig:ptllplot} and~\ref{fig:lamtungplot}. This implies that the $p_{T,ll}$ spectrum and the combination~$A_0 - A_2$ of angular coefficients provide observables that can help differentiate operators of the type~(\ref{fig:strom}) from the light-quark dipole interactions~(\ref{eq:LSMEFT}). We~further note that effective interactions of the form~(\ref{eq:vector}) affect the forward-backward asymmetry in $pp \to \gamma^\ast/Z +X \to l^+l^- + X$, as reflected in the angular coefficient $A_4$, and also influence the high-energy tails of the kinematic distributions in $pp \to Z + h \to l^+l^- + h$ production. These aspects are discussed, for example, in~\cite{Alioli:2018ljm,Alioli:2020kez} and~\cite{Gauld:2023gtb}, respectively. The first two papers also investigate the impact of semileptonic four-fermion operators and specific dimension-eight operators in DY processes. A discussion of such SMEFT effects is outside the scope of this article.

\section{Additional up-quark dipole results for $\bm{Z}$+jet production}
\label{app:more}

\begin{figure}[t]
\begin{center}
\includegraphics[scale=0.575]{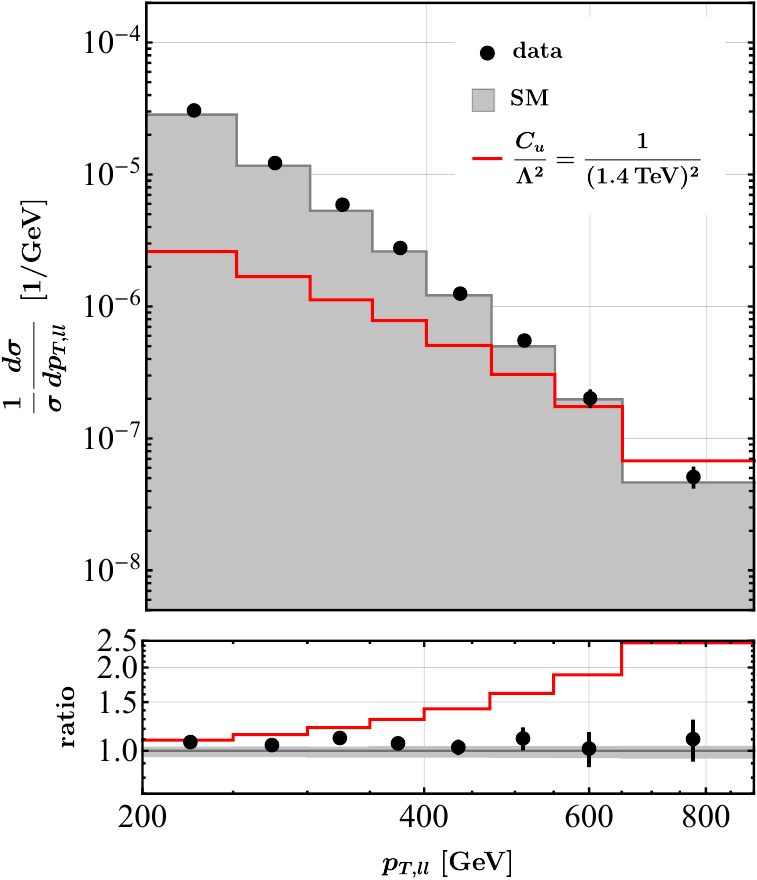} \qquad
\raisebox{10mm}[0mm][0mm]{\includegraphics[scale=0.575]{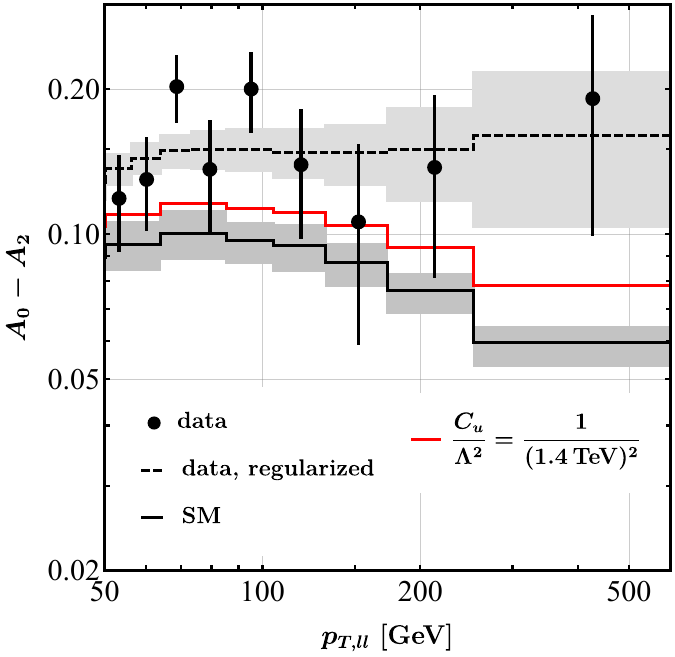}}
\end{center}
\vspace{-2mm} 
\caption{\label{fig:more} Left: Similar to~Figure~\ref{fig:ptllplot}. Right: Similar to Figure~\ref{fig:lamtungplot}. Both plots use the choice $C_u/\Lambda^2 = 1/(1.4 \, {\rm TeV})^2$. Further details can be found in the main text.}
\end{figure}

This appendix provides additional results for $Z$+jet production, employing the Wilson coefficient choice for the up-quark dipole operator from~\cite{Li:2024iyj}. Our predictions are shown in the two panels of~Figure~\ref{fig:more}. On the left side of the figure, we compare the ATLAS results~\cite{ATLAS:2019zci} with both the SM and a BSM prediction for the normalized~$p_{T,ll}$ spectrum in $Z$+jet production. The plot styles resemble that chosen in~Figure~\ref{fig:ptllplot}, and the displayed results were produced using the MC setup described in~Section~\ref{sec:MCsetup}. The Wilson coefficient choice $C_u/\Lambda^2= 1/(1.4 \, {\rm TeV})^2$ introduced in (\ref{eq:killphoton}) is identical to that of the up-quark dipole operator from~\cite{Li:2024iyj}. We observe that in the highest bin, $p_{T,ll} \in [650, 900] \, {\rm GeV}$, the chosen value of~$C_u/\Lambda^2$ leads to an enhancement of approximately $250\%$ compared to the SM prediction. Such large enhancements are firmly ruled out by the data. The red curve on the right-hand side in~Figure~\ref{fig:more}, illustrates the prediction for $A_0 - A_2$, again employing the value $C_u/\Lambda^2= 1/(1.4 \, {\rm TeV})^2$ for the relevant Wilson coefficient. The~plot styles used are identical to those in~Figure~\ref{fig:lamtungplot}, and the results were generated according to the description at the end of Section~\ref{sec:MCsetup}. The~plot demonstrates that including the considered up-quark dipole contribution noticeably affects the Lam-Tung relation, reducing the tension between theory and the ATLAS measurement~\cite{ATLAS:2016rnf} in the final bin with $p_{T,ll} \in [253, 600] \, {\rm GeV}$. However, it is important to note that even for $C_u/\Lambda^2= 1/(1.4 \, {\rm TeV})^2$, the BSM result for $A_0 - A_2$ is approximately $2\sigma$ ($2.5\sigma$) below the unregularized (regularized) data.

\end{appendix}


\begin{thebibliography}{10}
\providecommand{\url}[1]{\texttt{#1}}
\providecommand{\urlprefix}{URL }
\expandafter\ifx\csname urlstyle\endcsname\relax
 \providecommand{\doi}[1]{doi:\discretionary{}{}{}#1}\else
 \providecommand{\doi}{doi:\discretionary{}{}{}\begingroup
 \urlstyle{rm}\Url}\fi
\providecommand{\eprint}[2][]{\href{https://arxiv.org/abs/#2}{#2}}

\bibitem{ATLAS:2019zci}
G.~Aad \emph{et~al.},
\newblock \emph{{Measurement of the transverse momentum distribution of
 Drell\textendash{}Yan lepton pairs in proton\textendash{}proton collisions at
 $\sqrt{s}$~=~13~TeV with the ATLAS detector}},
\newblock Eur. Phys. J. C \textbf{80}(7), 616 (2020),
\newblock \doi{10.1140/epjc/s10052-020-8001-z},
\newblock \eprint{1912.02844}.

\bibitem{CMS:2022ubq}
A.~Tumasyan \emph{et~al.},
\newblock \emph{{Measurement of the mass dependence of the transverse momentum
 of lepton pairs in Drell-Yan production in proton-proton collisions at
 $\sqrt{s}$~=~13~TeV}},
\newblock Eur. Phys. J. C \textbf{83}(7), 628 (2023),
\newblock \doi{10.1140/epjc/s10052-023-11631-7},
\newblock \eprint{2205.04897}.

\bibitem{ATLAS:2023lhg}
G.~Aad \emph{et~al.},
\newblock \emph{{A precise determination of the strong-coupling constant from
 the recoil of $Z$ bosons with the ATLAS experiment at $\sqrt{s}$~=~8~TeV}}
 (2023),
\newblock \eprint{2309.12986}.

\bibitem{ATLAS:2024erm}
G.~Aad \emph{et~al.},
\newblock \emph{{Measurement of the W-boson mass and width with the ATLAS
 detector using proton-proton collisions at $\sqrt{s}$~=~7~TeV}} (2024),
\newblock \eprint{2403.15085}.

\bibitem{CMSMW}
CMS Collaboration,
\newblock \emph{{Measurement of the $W$ boson mass in proton-proton collisions at $\sqrt{s}$~=~13~TeV}} (2024),
\newblock \href{https://cds.cern.ch/record/2910372}{CMS-PAS-SMP-23-002}.

\bibitem{Buchmuller:1985jz}
W.~Buchm{\"u}ller and D.~Wyler,
\newblock \emph{{Effective Lagrangian Analysis of New Interactions and Flavor
 Conservation}},
\newblock Nucl. Phys. B \textbf{268}, 621 (1986),
\newblock \doi{10.1016/0550-3213(86)90262-2}.

\bibitem{Grzadkowski:2010es}
B.~Grzadkowski, M.~Iskrzynski, M.~Misiak and J.~Rosiek,
\newblock \emph{{Dimension-Six Terms in the Standard Model Lagrangian}},
\newblock JHEP \textbf{10}, 085 (2010),
\newblock \doi{10.1007/JHEP10(2010)085},
\newblock \eprint{1008.4884}.

\bibitem{Brivio:2017vri}
I.~Brivio and M.~Trott,
\newblock \emph{{The Standard Model as an Effective Field Theory}},
\newblock Phys. Rept. \textbf{793}, 1 (2019),
\newblock \doi{10.1016/j.physrep.2018.11.002},
\newblock \eprint{1706.08945}.

\bibitem{Isidori:2023pyp}
G.~Isidori, F.~Wilsch and D.~Wyler,
\newblock \emph{{The standard model effective field theory at work}},
\newblock Rev. Mod. Phys. \textbf{96}(1), 015006 (2024),
\newblock \doi{10.1103/RevModPhys.96.015006},
\newblock \eprint{2303.16922}.

\bibitem{Alioli:2018ljm}
S.~Alioli, W.~Dekens, M.~Girard and E.~Mereghetti,
\newblock \emph{{NLO QCD corrections to SM-EFT dilepton and electroweak Higgs
 boson production, matched to parton shower in POWHEG}},
\newblock JHEP \textbf{08}, 205 (2018),
\newblock \doi{10.1007/JHEP08(2018)205},
\newblock \eprint{1804.07407}.

\bibitem{Dawson:2018dxp}
S.~Dawson, P.~P. Giardino and A.~Ismail,
\newblock \emph{{Standard model EFT and the Drell-Yan process at high energy}},
\newblock Phys. Rev. D \textbf{99}(3), 035044 (2019),
\newblock \doi{10.1103/PhysRevD.99.035044},
\newblock \eprint{1811.12260}.

\bibitem{daSilvaAlmeida:2019cbr}
E.~da~Silva~Almeida, N.~Rosa-Agostinho, O.~J.~P. \'Eboli and M.~C.
 Gonzalez-Garcia,
\newblock \emph{{Light-quark dipole operators at the LHC}},
\newblock Phys. Rev. D \textbf{100}(1), 013003 (2019),
\newblock \doi{10.1103/PhysRevD.100.013003},
\newblock \eprint{1905.05187}.

\bibitem{Alioli:2020kez}
S.~Alioli, R.~Boughezal, E.~Mereghetti and F.~Petriello,
\newblock \emph{{Novel angular dependence in Drell-Yan lepton production via
 dimension-8 operators}},
\newblock Phys. Lett. B \textbf{809}, 135703 (2020),
\newblock \doi{10.1016/j.physletb.2020.135703},
\newblock \eprint{2003.11615}.

\bibitem{Horne:2020pot}
A.~Horne, J.~Pittman, M.~Snedeker, W.~Shepherd and J.~W. Walker,
\newblock \emph{{Shift-Type SMEFT Effects in Dileptons at the LHC}},
\newblock JHEP \textbf{03}, 118 (2021),
\newblock \doi{10.1007/JHEP03(2021)118},
\newblock \eprint{2007.12698}.

\bibitem{Torre:2020aiz}
R.~Torre, L.~Ricci and A.~Wulzer,
\newblock \emph{{On the W\&Y interpretation of high-energy Drell-Yan
 measurements}},
\newblock JHEP \textbf{02}, 144 (2021),
\newblock \doi{10.1007/JHEP02(2021)144},
\newblock \eprint{2008.12978}.

\bibitem{Greljo:2021kvv}
A.~Greljo, S.~Iranipour, Z.~Kassabov, M.~Madigan, J.~Moore, J.~Rojo, M.~Ubiali
 and C.~Voisey,
\newblock \emph{{Parton distributions in the SMEFT from high-energy Drell-Yan
 tails}},
\newblock JHEP \textbf{07}, 122 (2021),
\newblock \doi{10.1007/JHEP07(2021)122},
\newblock \eprint{2104.02723}.

\bibitem{Panico:2021vav}
G.~Panico, L.~Ricci and A.~Wulzer,
\newblock \emph{{High-energy EFT probes with fully differential Drell-Yan
 measurements}},
\newblock JHEP \textbf{07}, 086 (2021),
\newblock \doi{10.1007/JHEP07(2021)086},
\newblock \eprint{2103.10532}.

\bibitem{Dawson:2021ofa}
S.~Dawson and P.~P. Giardino,
\newblock \emph{{New physics through Drell-Yan standard model EFT measurements
 at NLO}},
\newblock Phys. Rev. D \textbf{104}(7), 073004 (2021),
\newblock \doi{10.1103/PhysRevD.104.073004},
\newblock \eprint{2105.05852}.

\bibitem{Boughezal:2022nof}
R.~Boughezal, Y.~Huang and F.~Petriello,
\newblock \emph{{Exploring the SMEFT at dimension eight with Drell-Yan
 transverse momentum measurements}},
\newblock Phys. Rev. D \textbf{106}(3), 036020 (2022),
\newblock \doi{10.1103/PhysRevD.106.036020},
\newblock \eprint{2207.01703}.

\bibitem{Allwicher:2022mcg}
L.~Allwicher, D.~A. Faroughy, F.~Jaffredo, O.~Sumensari and F.~Wilsch,
\newblock \emph{{HighPT: A tool for~ high-$p_T$ Drell-Yan tails beyond the
 standard model}},
\newblock Comput. Phys. Commun. \textbf{289}, 108749 (2023),
\newblock \doi{10.1016/j.cpc.2023.108749},
\newblock \eprint{2207.10756}.

\bibitem{Boughezal:2023nhe}
R.~Boughezal, Y.~Huang and F.~Petriello,
\newblock \emph{{Impact of high invariant-mass Drell-Yan forward-backward
 asymmetry measurements on SMEFT fits}},
\newblock Phys. Rev. D \textbf{108}(7), 076008 (2023),
\newblock \doi{10.1103/PhysRevD.108.076008},
\newblock \eprint{2303.08257}.

\bibitem{Li:2024iyj}
X.~Li, B.~Yan and C.~P. Yuan,
\newblock \emph{{Lam-Tung relation breaking in $Z$ boson production as a probe
 of SMEFT effects}} (2024),
\newblock \eprint{2405.04069}.

\bibitem{Lam:1978zr}
C.~S. Lam and W.-K. Tung,
\newblock \emph{{Structure Function Relations at Large Transverse Momenta in
 Lepton Pair Production Processes}},
\newblock Phys. Lett. B \textbf{80}, 228 (1979),
\newblock \doi{10.1016/0370-2693(79)90204-1}.

\bibitem{Lam:1978pu}
C.~S. Lam and W.-K. Tung,
\newblock \emph{{A Systematic Approach to Inclusive Lepton Pair Production in
 Hadronic Collisions}},
\newblock Phys. Rev. D \textbf{18}, 2447 (1978),
\newblock \doi{10.1103/PhysRevD.18.2447}.

\bibitem{Lam:1980uc}
C.~S. Lam and W.-K. Tung,
\newblock \emph{{A Parton Model Relation Without QCD Modifications in Lepton
 Pair Productions}},
\newblock Phys. Rev. D \textbf{21}, 2712 (1980),
\newblock \doi{10.1103/PhysRevD.21.2712}.

\bibitem{Kley:2021yhn}
J.~Kley, T.~Theil, E.~Venturini and A.~Weiler,
\newblock \emph{{Electric dipole moments at one-loop in the dimension-6
 SMEFT}},
\newblock Eur. Phys. J. C \textbf{82}(10), 926 (2022),
\newblock \doi{10.1140/epjc/s10052-022-10861-5},
\newblock \eprint{2109.15085}.

\bibitem{ParticleDataGroup:2024cfk}
S.~Navas \emph{et~al.},
\newblock \emph{{Review of particle physics}},
\newblock Phys. Rev. D \textbf{110}(3), 030001 (2024),
\newblock \doi{10.1103/PhysRevD.110.030001}.

\bibitem{ALEPH:2005ab}
S.~Schael \emph{et~al.},
\newblock \emph{{Precision electroweak measurements on the $Z$ resonance}},
\newblock Phys. Rept. \textbf{427}, 257 (2006),
\newblock \doi{10.1016/j.physrep.2005.12.006},
\newblock \eprint{hep-ex/0509008}.

\bibitem{Farina:2016rws}
M.~Farina, G.~Panico, D.~Pappadopulo, J.~T. Ruderman, R.~Torre and A.~Wulzer,
\newblock \emph{{Energy helps accuracy: electroweak precision tests at hadron
 colliders}},
\newblock Phys. Lett. B \textbf{772}, 210 (2017),
\newblock \doi{10.1016/j.physletb.2017.06.043},
\newblock \eprint{1609.08157}.

\bibitem{Greljo:2017vvb}
A.~Greljo and D.~Marzocca,
\newblock \emph{{High-$p_T$ dilepton tails and flavor physics}},
\newblock Eur. Phys. J. C \textbf{77}(8), 548 (2017),
\newblock \doi{10.1140/epjc/s10052-017-5119-8},
\newblock \eprint{1704.09015}.

\bibitem{Alioli:2017jdo}
S.~Alioli, M.~Farina, D.~Pappadopulo and J.~T. Ruderman,
\newblock \emph{{Precision Probes of QCD at High Energies}},
\newblock JHEP \textbf{07}, 097 (2017),
\newblock \doi{10.1007/JHEP07(2017)097},
\newblock \eprint{1706.03068}.

\bibitem{Alioli:2017nzr}
S.~Alioli, M.~Farina, D.~Pappadopulo and J.~T. Ruderman,
\newblock \emph{{Catching a New Force by the Tail}},
\newblock Phys. Rev. Lett. \textbf{120}(10), 101801 (2018),
\newblock \doi{10.1103/PhysRevLett.120.101801},
\newblock \eprint{1712.02347}.

\bibitem{Banerjee:2018bio}
S.~Banerjee, C.~Englert, R.~S. Gupta and M.~Spannowsky,
\newblock \emph{{Probing Electroweak Precision Physics via boosted
 Higgs-strahlung at the LHC}},
\newblock Phys. Rev. D \textbf{98}(9), 095012 (2018),
\newblock \doi{10.1103/PhysRevD.98.095012},
\newblock \eprint{1807.01796}.

\bibitem{Grojean:2018dqj}
C.~Grojean, M.~Montull and M.~Riembau,
\newblock \emph{{Diboson at the LHC vs LEP}},
\newblock JHEP \textbf{03}, 020 (2019),
\newblock \doi{10.1007/JHEP03(2019)020},
\newblock \eprint{1810.05149}.

\bibitem{DiLuzio:2018jwd}
L.~Di~Luzio, R.~Gr\"ober and G.~Panico,
\newblock \emph{{Probing new electroweak states via precision measurements at
 the LHC and future colliders}},
\newblock JHEP \textbf{01}, 011 (2019),
\newblock \doi{10.1007/JHEP01(2019)011},
\newblock \eprint{1810.10993}.

\bibitem{Fuentes-Martin:2020lea}
J.~Fuentes-Martin, A.~Greljo, J.~Martin~Camalich and J.~D. Ruiz-Alvarez,
\newblock \emph{{Charm physics confronts high-p$_{T}$ lepton tails}},
\newblock JHEP \textbf{11}, 080 (2020),
\newblock \doi{10.1007/JHEP11(2020)080},
\newblock \eprint{2003.12421}.

\bibitem{Haisch:2021hcg}
U.~Haisch and G.~Koole,
\newblock \emph{{Beautiful and charming chromodipole moments}},
\newblock JHEP \textbf{09}, 133 (2021),
\newblock \doi{10.1007/JHEP09(2021)133},
\newblock \eprint{2106.01289}.

\bibitem{Haisch:2023upo}
U.~Haisch, L.~Schnell and J.~Weiss,
\newblock \emph{{LHC tau-pair production constraints on $a_\tau$ and
 $d_\tau$}},
\newblock SciPost Phys. \textbf{16}(2), 048 (2024),
\newblock \doi{10.21468/SciPostPhys.16.2.048},
\newblock \eprint{2307.14133}.

\bibitem{Gauld:2023gtb}
R.~Gauld, U.~Haisch and L.~Schnell,
\newblock \emph{{SMEFT at NNLO+PS: $Vh$ production}},
\newblock JHEP \textbf{01}, 192 (2024),
\newblock \doi{10.1007/JHEP01(2024)192},
\newblock \eprint{2311.06107}.

\bibitem{Hiller:2024vtr}
G.~Hiller and D.~Wendler,
\newblock \emph{{Missing energy plus jet in the SMEFT}},
\newblock JHEP \textbf{09}, 009 (2024),
\newblock \doi{10.1007/JHEP09(2024)009},
\newblock \eprint{2403.17063}.

\bibitem{H1:2009pze}
F.~D. Aaron \emph{et~al.},
\newblock \emph{{Combined Measurement and QCD Analysis of the Inclusive $e^\pm p$
 Scattering Cross Sections at HERA}},
\newblock JHEP \textbf{01}, 109 (2010),
\newblock \doi{10.1007/JHEP01(2010)109},
\newblock \eprint{0911.0884}.

\bibitem{Callan:1969uq}
C.~G. Callan, Jr. and D.~J. Gross,
\newblock \emph{{High-energy electroproduction and the constitution of the
 electric current}},
\newblock Phys. Rev. Lett. \textbf{22}, 156 (1969),
\newblock \doi{10.1103/PhysRevLett.22.156}.

\bibitem{Buchmuller:1987ur}
W.~Buchm{\"u}ller, B.~Lampe and N.~Vlachos,
\newblock \emph{{Contact Interactions and the Callan-Gross Relation}},
\newblock Phys. Lett. B \textbf{197}, 379 (1987),
\newblock \doi{10.1016/0370-2693(87)90404-7}.

\bibitem{Collins:1977iv}
J.~C. Collins and D.~E. Soper,
\newblock \emph{{Angular Distribution of Dileptons in High-Energy Hadron
 Collisions}},
\newblock Phys. Rev. D \textbf{16}, 2219 (1977),
\newblock \doi{10.1103/PhysRevD.16.2219}.

\bibitem{Arteaga-Romero:1983llb}
N.~Arteaga-Romero, A.~Nicolaidis and J.~Silva,
\newblock \emph{{$Z^0$ Production at the $p \bar{p}$ Collider and the Spin of
 the Gluon}},
\newblock Phys. Rev. Lett. \textbf{52}, 172 (1984),
\newblock \doi{10.1103/PhysRevLett.52.172}.

\bibitem{Mirkes:1994dp}
E.~Mirkes and J.~Ohnemus,
\newblock \emph{{Angular distributions of Drell-Yan lepton pairs at the
 Tevatron: Order $\alpha_s^{2}$ corrections and Monte Carlo studies}},
\newblock Phys. Rev. D \textbf{51}, 4891 (1995),
\newblock \doi{10.1103/PhysRevD.51.4891},
\newblock \eprint{hep-ph/9412289}.

\bibitem{Gehrmann-DeRidder:2015wbt}
A.~Gehrmann-De~Ridder, T.~Gehrmann, E.~W.~N. Glover, A.~Huss and T.~A. Morgan,
\newblock \emph{{Precise QCD predictions for the production of a Z boson in
 association with a hadronic jet}},
\newblock Phys. Rev. Lett. \textbf{117}(2), 022001 (2016),
\newblock \doi{10.1103/PhysRevLett.117.022001},
\newblock \eprint{1507.02850}.

\bibitem{Gehrmann-DeRidder:2016jns}
A.~Gehrmann-De~Ridder, T.~Gehrmann, E.~W.~N. Glover, A.~Huss and T.~A. Morgan,
\newblock \emph{{NNLO QCD corrections for Drell-Yan $p_T^Z$ and $\phi^*$
 observables at the LHC}},
\newblock JHEP \textbf{11}, 094 (2016),
\newblock \doi{10.1007/JHEP11(2016)094},
\newblock [Erratum: JHEP \textbf{10}, 126 (2018)],
\newblock \eprint{1610.01843}.

\bibitem{Denner:2011vu}
A.~Denner, S.~Dittmaier, T.~Kasprzik and A.~M{\"u}ck,
\newblock \emph{{Electroweak corrections to dilepton + jet production at hadron
 colliders}},
\newblock JHEP \textbf{06}, 069 (2011),
\newblock \doi{10.1007/JHEP06(2011)069},
\newblock \eprint{1103.0914}.

\bibitem{Alloul:2013bka}
A.~Alloul, N.~D. Christensen, C.~Degrande, C.~Duhr and B.~Fuks,
\newblock \emph{{FeynRules 2.0 - A complete toolbox for tree-level
 phenomenology}},
\newblock Comput. Phys. Commun. \textbf{185}, 2250 (2014),
\newblock \doi{10.1016/j.cpc.2014.04.012},
\newblock \eprint{1310.1921}.

\bibitem{Degrande:2011ua}
C.~Degrande, C.~Duhr, B.~Fuks, D.~Grellscheid, O.~Mattelaer and T.~Reiter,
\newblock \emph{{UFO - The Universal FeynRules Output}},
\newblock Comput. Phys. Commun. \textbf{183}, 1201 (2012),
\newblock \doi{10.1016/j.cpc.2012.01.022},
\newblock \eprint{1108.2040}.

\bibitem{Alwall:2014hca}
J.~Alwall, R.~Frederix, S.~Frixione, V.~Hirschi, F.~Maltoni, O.~Mattelaer,
 H.~S. Shao, T.~Stelzer, P.~Torrielli and M.~Zaro,
\newblock \emph{{The automated computation of tree-level and next-to-leading
 order differential cross sections, and their matching to parton shower
 simulations}},
\newblock JHEP \textbf{07}, 079 (2014),
\newblock \doi{10.1007/JHEP07(2014)079},
\newblock \eprint{1405.0301}.

\bibitem{Sjostrand:2014zea}
T.~Sj\"ostrand, S.~Ask, J.~R. Christiansen, R.~Corke, N.~Desai, P.~Ilten,
 S.~Mrenna, S.~Prestel, C.~O. Rasmussen and P.~Z. Skands,
\newblock \emph{{An introduction to PYTHIA 8.2}},
\newblock Comput. Phys. Commun. \textbf{191}, 159 (2015),
\newblock \doi{10.1016/j.cpc.2015.01.024},
\newblock \eprint{1410.3012}.

\bibitem{NNPDF:2017mvq}
R.~D. Ball \emph{et~al.},
\newblock \emph{{Parton distributions from high-precision collider data}},
\newblock Eur. Phys. J. C \textbf{77}(10), 663 (2017),
\newblock \doi{10.1140/epjc/s10052-017-5199-5},
\newblock \eprint{1706.00428}.

\bibitem{ATLAS:2016rnf}
G.~Aad \emph{et~al.},
\newblock \emph{{Measurement of the angular coefficients in $Z$-boson events
 using electron and muon pairs from data taken at $\sqrt{s}$~=~8~TeV with the
 ATLAS detector}},
\newblock JHEP \textbf{08}, 159 (2016),
\newblock \doi{10.1007/JHEP08(2016)159},
\newblock \eprint{1606.00689}.

\bibitem{CMS:2015cyj}
V.~Khachatryan \emph{et~al.},
\newblock \emph{{Angular coefficients of $Z$ bosons produced in pp collisions at
 $\sqrt{s}$~=~8~TeV and decaying to $\mu^+ \mu^-$ as a function of transverse
 momentum and rapidity}},
\newblock Phys. Lett. B \textbf{750}, 154 (2015),
\newblock \doi{10.1016/j.physletb.2015.08.061},
\newblock \eprint{1504.03512}.

\bibitem{Gauld:2017tww}
R.~Gauld, A.~Gehrmann-De~Ridder, T.~Gehrmann, E.~W.~N. Glover and A.~Huss,
\newblock \emph{{Precise predictions for the angular coefficients in $Z$-boson
 production at the LHC}},
\newblock JHEP \textbf{11}, 003 (2017),
\newblock \doi{10.1007/JHEP11(2017)003},
\newblock \eprint{1708.00008}.

\bibitem{Butterworth:2015oua}
J.~Butterworth \emph{et~al.},
\newblock \emph{{PDF4LHC recommendations for LHC Run II}},
\newblock J. Phys. G \textbf{43}, 023001 (2016),
\newblock \doi{10.1088/0954-3899/43/2/023001},
\newblock \eprint{1510.03865}.

\bibitem{Hahn:2000kx}
T.~Hahn,
\newblock \emph{{Generating Feynman diagrams and amplitudes with FeynArts 3}},
\newblock Comput. Phys. Commun. \textbf{140}, 418 (2001),
\newblock \doi{10.1016/S0010-4655(01)00290-9},
\newblock \eprint{hep-ph/0012260}.

\bibitem{Hahn:2016ebn}
T.~Hahn, S.~Pa\ss{}ehr and C.~Schappacher,
\newblock \emph{{FormCalc 9 and Extensions}},
\newblock PoS \textbf{LL2016}, 068 (2016),
\newblock \doi{10.1088/1742-6596/762/1/012065},
\newblock \eprint{1604.04611}.

\bibitem{Gauld:2021zmq}
R.~Gauld,
\newblock \emph{{A massive variable flavour number scheme for the Drell-Yan
 process}},
\newblock SciPost Phys. \textbf{12}(1), 024 (2022),
\newblock \doi{10.21468/SciPostPhys.12.1.024},
\newblock \eprint{2107.01226}.

\bibitem{Cowan:2010js}
G.~Cowan, K.~Cranmer, E.~Gross and O.~Vitells,
\newblock \emph{{Asymptotic formulae for likelihood-based tests of new
 physics}},
\newblock Eur. Phys. J. C \textbf{71}, 1554 (2011),
\newblock \doi{10.1140/epjc/s10052-011-1554-0},
\newblock [Erratum: Eur. Phys. J. C \textbf{73}, 2501 (2013)],
\newblock \eprint{1007.1727}.

\bibitem{Lai:2010nw}
H.-L. Lai, J.~Huston, Z.~Li, P.~Nadolsky, J.~Pumplin, D.~Stump and C.~P. Yuan,
\newblock \emph{{Uncertainty induced by QCD coupling in the CTEQ global
 analysis of parton distributions}},
\newblock Phys. Rev. D \textbf{82}, 054021 (2010),
\newblock \doi{10.1103/PhysRevD.82.054021},
\newblock \eprint{1004.4624}.

\bibitem{Lindert:2017olm}
J.~M. Lindert \emph{et~al.},
\newblock \emph{{Precise predictions for $V$+~jets dark matter backgrounds}},
\newblock Eur. Phys. J. C \textbf{77}(12), 829 (2017),
\newblock \doi{10.1140/epjc/s10052-017-5389-1},
\newblock \eprint{1705.04664}.

\end{thebibliography}



\end{fmffile}
\end{document}